\documentclass{jfm}
\usepackage{graphicx}
\usepackage{epstopdf, epsfig}
\usepackage{caption,subcaption}
\usepackage{color}
\usepackage{amsmath}
\usepackage{dcolumn}
\usepackage{bm}

\usepackage{array}

\usepackage[utf8]{inputenc}
\usepackage[T1]{fontenc}
\usepackage{ae,aecompl}

\usepackage{color}
\definecolor{darkblue}{rgb}{0,0,0.6}
\definecolor{darkred}{rgb}{0.6,0,0}
\definecolor{darkgreen}{rgb}{0,0.6,0}

\usepackage[colorlinks=true,urlcolor=darkblue,citecolor=darkblue,linkcolor=darkred,hyperfootnotes=false]{hyperref}

\newcommand{\bv}[1]{{\boldsymbol #1}}

\newcommand{\Reyc}{{\rm Re}_{\rm c}}
\newcommand{\Reynul}{\rm Re}

\shorttitle{Do extreme events trigger turbulence decay?}
\shortauthor{T. Nemoto and A. Alexakis}

\title{Do extreme events trigger turbulence decay? --
a numerical study of turbulence decay time 
 in pipe flows}

\author{Takahiro Nemoto \aff{1,2}
 \and Alexandros Alexakis\aff{3}}

\affiliation{\aff{1} Philippe Meyer Institute for Theoretical Physics, Physics Department, \'Ecole Normale Sup\'erieure \& PSL Research University, 24 rue Lhomond, 75231 Paris Cedex 05, France
\aff{2} Mathematical Modelling of Infectious Diseases Unit, Institut Pasteur, 25-28 Rue du Docteur Roux, 75015 Paris, France
\aff{3} Laboratoire de Physique de l'Ecole normale sup\'erieure, ENS, Universit\'e PSL, CNRS, Sorbonne Universit\'e, Universit\'e Paris-Diderot, Sorbonne Paris Cit\'e, Paris, France}

\begin{document}

\maketitle

\begin{abstract}

Turbulence locally created in laminar pipe flows shows sudden decay or splitting after a stochastic waiting time. In laboratory experiments, the mean waiting time was observed to increase double-exponentially as the Reynolds number (Re) approaches its critical value. To understand the origin of this double-exponential increase, we perform many pipe flow direct numerical simulations of the Navier-Stokes equations, and measure the cumulative histogram of the maximum axial vorticity field over the pipe (turbulence intensity). 
In the domain where the turbulence intensity is not small, we observe that the histogram is well-approximated by  the Gumbel extreme-value distribution. The smallest turbulence intensity in this domain roughly corresponds to the transition value between the locally stable turbulence and a meta-stable (edge) state. 
Studying the Re dependence of the fitting parameters in this distribution, we derive that the time scale of the transition between these two states increases double-exponentially as Re approaches its critical value. On the contrary, in smaller turbulence intensities below this domain, we observe that the distribution is not sensible to the change of Re. This means that the decay time of the meta-stable state (to the laminar state) is stochastic but Re-independent in average. 
Our observation suggests that the conjecture made by Goldenfeld {\it et al.} to derive the double-exponential increase of turbulence decay time is approximately satisfied in the range of Re we studied. 
We also discuss using another extreme-value distribution, Fr\'echet distribution, instead of the Gumbel distribution to approximate the histogram of the turbulence intensify, which reveals interesting  perspectives.

\end{abstract}  

\begin{keywords} Turbulence, turbulent-laminar transitions, pipe flows, extreme value statistics
\end{keywords}

\section{Introduction}

The laminar to turbulent transition in pipe flow is one of the most important problems in fluid mechanics \citep{doi:10.1146/annurev.fluid.39.050905.110308}, initiated by O. Reynolds in 1887 \citep{reynolds1883experimental}. The flow in a pipe is characterised by a non-dimensional parameter, the Reynolds number defined as ${\Reynul}=U R /\nu$ where $U$ is the flow velocity, $R$ the diameter of the pipe, and $\nu$ the dynamic viscosity of the fluid. The question about the critical Reynolds number $\Reyc$ is the following: At which Reynolds number does the flow in a pipe become turbulent from laminar (and {\it vice versa})? Though its apparent simplicity, answering this question was not straightforward due to a number of technical reasons (see \citep{Eckhardt449} for a historical review of the critical Reynolds number). After the 2000s, a localised turbulent state (Fig.~\ref{fig:structure}) called ``puff'' \citep{wygnanski1973transition} created by a local perturbation to laminar flows has been studied in detail. The puff shows a sudden decay or splitting after a stochastic waiting time, following a memoryless exponential distribution \citep{hof2006finite,PhysRevLett.101.214501,de_Lozar589,avila2010transient,kuik2010quantitative}. These studies culminated in the estimation of the critical Reynolds number in 2011 \citep{Avila192}, where $\Reyc$ was determined as the ${\rm Re}$ at which the two typical times of the decay and splitting  become equal. The obtained $\Reyc$ was about 2040~\citep{Avila192}.

The critical Reynolds number was determined in laboratory experiments using long pipes. In direct numerical simulations (DNS) of the Navier-Stokes equations, the observation of the critical Reynolds number has not yet been achieved because of too high computational costs. The current state of the art is to measure decay events up to $\Reynul=1900$ and splitting events down to $2100$ \citep{Avila192}. There are several important benefits in measuring these decay and splitting events in DNS. First, in experiments, unknown background noise that affects the results could exist. See, for example,  \citep{Eckhardt449} for a history of the struggle to determine the upper critical Reynolds number due to small background fluctuations. In DNS on the other hand, we know all the origins of the artificial noise, such as insufficient mesh sizes or periodic boundary effects, which can be easily controlled. Second, the quantities that are measurable in experiments are limited. For example, to determine the critical Reynolds number, a double-exponential fitting curve was heuristically used \citep{hof2006finite,PhysRevLett.101.214501,Avila192}. The origin of this double-exponential law was discussed using the extreme value theory \citep{PhysRevE.81.035304,Goldenfeld2017}, where Goldenfeld {\it et al.} conjectured that the cumulative distribution function (CDF) of maximum kinetic energy fluctuations has a certain form.
Measurements of this probability function are out of reach in experiments, but can be achieved in DNS.

In this article by using high performance computing resources described in Acknowledgment, we study the statistics of turbulence decay in pipe flows with DNS. We perform more than 1000 independent pipe flow simulations in parallel (see Appendix~\ref{appA}) and determine the decay time up to $\Reynul=2000$. Especially, our aim is to investigate if the CDF of a maximum turbulence intensity is described by the double-exponential Gumbel distribution as conjectured  by Goldenfeld {\it et al.} \citep{PhysRevE.81.035304,Goldenfeld2017} to derive the double-exponential law.

\section{Set-up}

\begin{figure}
\begin{center}
\includegraphics[width=120mm]{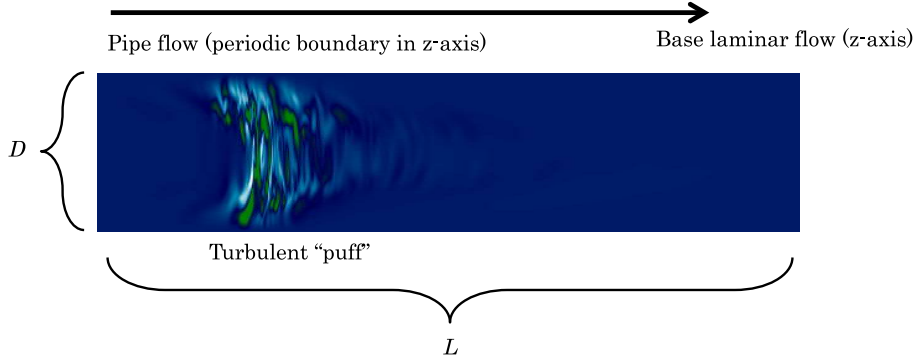}
\caption{\label{fig:structure}  
A pipe flow simulation with a diameter $D$ and length $L$, containing a single turbulent puff, is considered. In the figure, the velocity component perpendicular to the figure is visualised using a colourmap. We perform many  pipe flow simulations during a certain time interval, where the initial conditions are slightly different from each other (so that the dynamics are independent after a relaxation time). We consider the sum of all the durations of the simulations (where the  initial relaxation intervals and the intervals after the puff decays are subtracted).  We denote this total duration by  $T$ and also the total number of decay events observed in all the simulations by  $n_{\rm d}$. From this pair of quantities, we estimate the typical decay time $\tau_{\rm d}$ as $ T/n_{\rm d}$ and its error bars by using Bayesian inference as detailed in the main text. See also Appendix \ref{appA} for more details of the simulation architecture.}
\end{center}
\end{figure}

We consider three-dimensional pipe flows (with pipe length $L$ and pipe diameter $D$) where the boundaries are periodic for the z-axis and no-slip along the pipe (Fig.~\ref{fig:structure}). The mean flow speed in the z direction is denoted by $U_{\rm b}$ and we use the basic unit of length and time as $D$ and $D/U_{\rm b}$ throughout this paper. The velocity field is denoted by ${\bv u}(\bv r,t)$ and we simulate its evolution by solving the Navier-Stokes equations using an open source code (openpipeflow~\citep{WILLIS2017124}) whose validity has been widely tested in many works (see Appendix~\ref{appA} for the simulation detail). 
We simulate pipe flows for the Reynolds numbers below the critical value ${\Reynul}<\Reyc\sim 2040$, where the flows tend to be laminar. We start the simulations with initial conditions where localised turbulence, {\it i.e.}, a turbulent puff, exists (see Appendix~\ref{appA} for more details). If the Reynolds number is not too small (say, ${\Reynul}>1850$), localised turbulent dynamics are sustained (Fig.~\ref{fig:structure}). It quickly forgets their initial conditions and eventually decays after the stochastic time $t$, following an exponential distribution function \citep{hof2006finite,PhysRevLett.101.214501,de_Lozar589,avila2010transient,kuik2010quantitative}: $p_{\rm d}(t)=(1/\tau_{\rm d})\exp(-t/\tau_{\rm d})$, where $\tau_{\rm d}$ is the typical decay time. This time scale $\tau_{\rm d}$ is our target.

\section{Results}

\subsection{Measurements of $\tau_{\rm d}$ }
\label{seq:measurement_of_tau}

To measure $\tau_{\rm d}$, we perform many pipe flow simulations 
and measure the total time $T$ of all the simulations (ignoring initial relaxation intervals) and the number of decay events $n_{\rm d}$ that we observe during all the simulations. 
The typical decay time $\tau_{\rm d}$ can be estimated as $T/n_{\rm d}.$ This estimate converges to the true value as we observe more decay events.
 Since we observe only a few decay events when the Reynolds number is large (as shown in Table~\ref{Table:M_and_K}),  it is then important to calculate accurately the statistical errors of this estimate. To this end, we use Bayesian inference as detailed as follows: Since the decay time is distributed exponentially \citep{hof2006finite,PhysRevLett.101.214501,de_Lozar589,avila2010transient,kuik2010quantitative}, the probability of observing 
 $n_{\rm d}$ decay events during a time interval $T$ follows the Poisson distribution 
\begin{equation}
p_{\rm poisson}(n_{\rm d})=\frac{e^{-\lambda_{\rm d} T}}{n_{\rm d}!}
\left (\lambda_{\rm d} T \right )^{n_{\rm d}},
\label{eq:poissondist}
\end{equation}
where $\lambda_{\rm d}\equiv 1/\tau_{\rm d}$.  To derive a probability distribution of the rate $\lambda_{\rm d}$ from this Poisson distribution, we use Bayes' rule. It allows us to construct a posterior probability distribution $p_{\rm posterior}(\lambda_{\rm d}|n_d)$, {\it i.e.}, the probability distribution of the parameter $\lambda_{\rm d}$ for a given observed data $n_d$, as follows: 
\begin{equation}
P_{\rm posterior}(\lambda_{\rm d}|n_d) \propto p_{\rm poisson}(n_{\rm d}) Q_{\rm prior}(\lambda_{\rm d}). 
\label{eq:posterior}
\end{equation}
Here $Q_{\rm prior}(\lambda_{\rm d})$ is a prior probability distribution that represents the initial guess of the parameter distribution. In our case, as we do not know the probability of $\lambda_{\rm d}$ {\it a priori}, we use a Jeffreys prior (an uninformative prior) defined as the square root of the determinant of the Fisher information \citep{box2011bayesian}, $Q_{\rm prior}(\lambda_{\rm d}) \propto 1/\sqrt{\lambda_{\rm d}}$. 
With this posterior distribution $P_{\rm posterior}(\lambda_{\rm d}|n_d)$, we define the error bars using 95\% confidence intervals: we define 2.5th percentile $\lambda_{{\rm d}}^{2.5}$ and 97.5th percentile $\lambda_{{\rm d}}^{97.5}$ as the values of $\lambda_{\rm d}$ at which the cumulative posterior probability distribution takes 0.025 and 0.975, respectively, {\it i.e.,}  $\int_{0}^{\lambda_{{\rm d}}^{2.5}} d\lambda_{\rm d} P_{\rm posterior}(\lambda_{\rm d}|n_d) = 0.025$ and $\int_{0}^{\lambda_{{\rm d}}^{97.5}} d\lambda_{\rm d} P_{\rm posterior}(\lambda_{\rm d}|n_d) = 0.975$. Using these percentiles,  the 95\% confidence interval is then defined as 
$\lambda_{{\rm d}}^{2.5}<\lambda_{\rm d}<\lambda_{{\rm d}}^{97.5}$.

\begin{figure}
\begin{center}
\includegraphics[width=100mm]{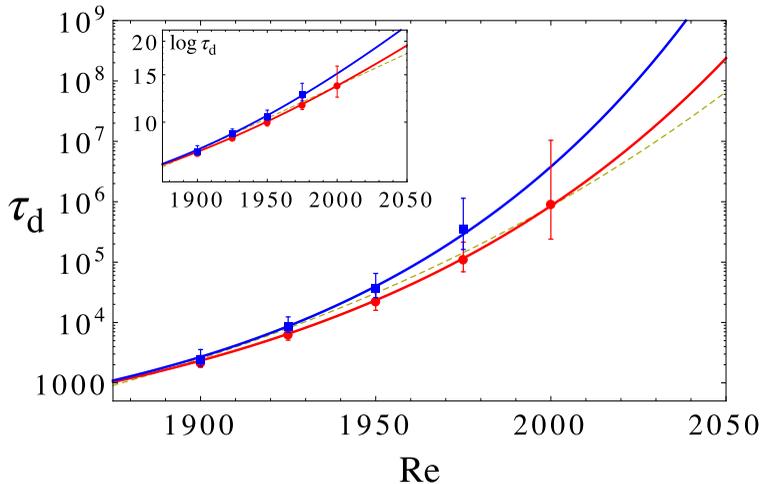}
\caption{\label{fig:decaytime}
The typical decay time $\tau_{\rm d}$ of turbulent puff in pipe flows obtained from DNS (red dots for $L=50D$ and blue squared for $L=100D$). Double exponential curves fitted to experiments \citep{Avila192} (yellow dashed line) and our theoretical lines (\ref{eq:tau_decay_last}) are also plotted as red and blue solid lines. In the theoretical line (\ref{eq:tau_decay_last}), the parameters $a,b$ are determined in Fig.~\ref{fig:cumumaximum_c}(b), $\delta t$ is set to $1/4$, and $\Pi(h_{\rm x})$ is measured in Fig.~\ref{fig:cumumaximum_a}(b). The error bars show 95\% confidence interval (see the main text for the definition). The inset shows the same data but with $\log \tau_{\rm d}$.
}
\end{center}
\end{figure}

We show in Fig.~\ref{fig:decaytime} the obtained $\tau_{\rm d}$ for different pipe lengths $L=50D, L=100D$ (red circles and blue squares, respectively) together with the experimentally fitted double exponential (yellow dashed) curve used in \citep{Avila192}. For the Reynolds number up to 1900 (which has been studied so far using DNS), $\tau_{\rm d}$ does not depend on the pipe length, and the results for both pipe lengths agree very well with the experimental data. However, as the Reynolds number increases, we observe that the results for $L=100D$ deviate from those for $L=50D$ and for the experiments. In DNS, periodic boundary conditions introduce  confinement effects on the puff:  An insufficient pipe length in DNS  prevents the puff from fully developing as it would in an infinite pipe. In the range of the pipe length we studied, this confinement facilitates decay events ({\it i.e.}, as the pipe length increases, the average decay time also increases). Note that the experimental results are of the same order as the DNS result for $L=50D$, even though the pipe lengths used for the experiments were much longer \citep{Avila192}.  Further numerical studies are  necessary to understand the convergence of the decay time as the pipe length increases. Below we discuss the derivation of the double-exponential formula for each {\it fixed} $L=50D$ and $L=100D$.

\subsection{Statistical property of the maximum turbulence intensity}
\label{sec:maximum_turbulence}
For  the Reynolds number dependence of the time scale $\tau_{\rm d}$, 
a double-exponential fitting curve $\exp \left[ \exp (\alpha {\Reynul} + \beta)\right ]$ (with fitting parameters $\alpha,\beta$) was heuristically used \citep{Avila192}. Although this function can fit well to the experimentally observed time scale $\tau_{\rm d}$, the origin of this double exponential form is still conjectural. 
In the conjecture made by Goldenfeld {\it et al.} in 2010 \citep{PhysRevE.81.035304}, they assumed that the maximum of the kinetic energy fluctuations over the pipe is distributed double exponentially (the Gumbel distribution function) due to the extreme value theory \citep{fisher1928limiting,gumbel1935valeurs}. When this maximum goes below a certain threshold, turbulence decays. Assuming the linear dependence on  $\Reynul$ of the parameters in the Gumbel distribution function, they thus derived the double-exponential increase of the time scale $\tau_{\rm d}$. Mathematically proving the validity of the extreme value theory and the linear scaling of the fitting parameters seems impossible, thus verifications in experiments or numerical simulations are needed. In laboratory experiments, the verification of this extreme value theory is not easy, as obtaining the maximum of a velocity field within a tiny turbulent puff is a non-trivial procedure. In this respect, numerical simulations have a strong advantage, because velocity fields are precisely tractable and the maximum of the fields is well-defined.

\begin{figure}
\begin{center}
\includegraphics[width=67mm]{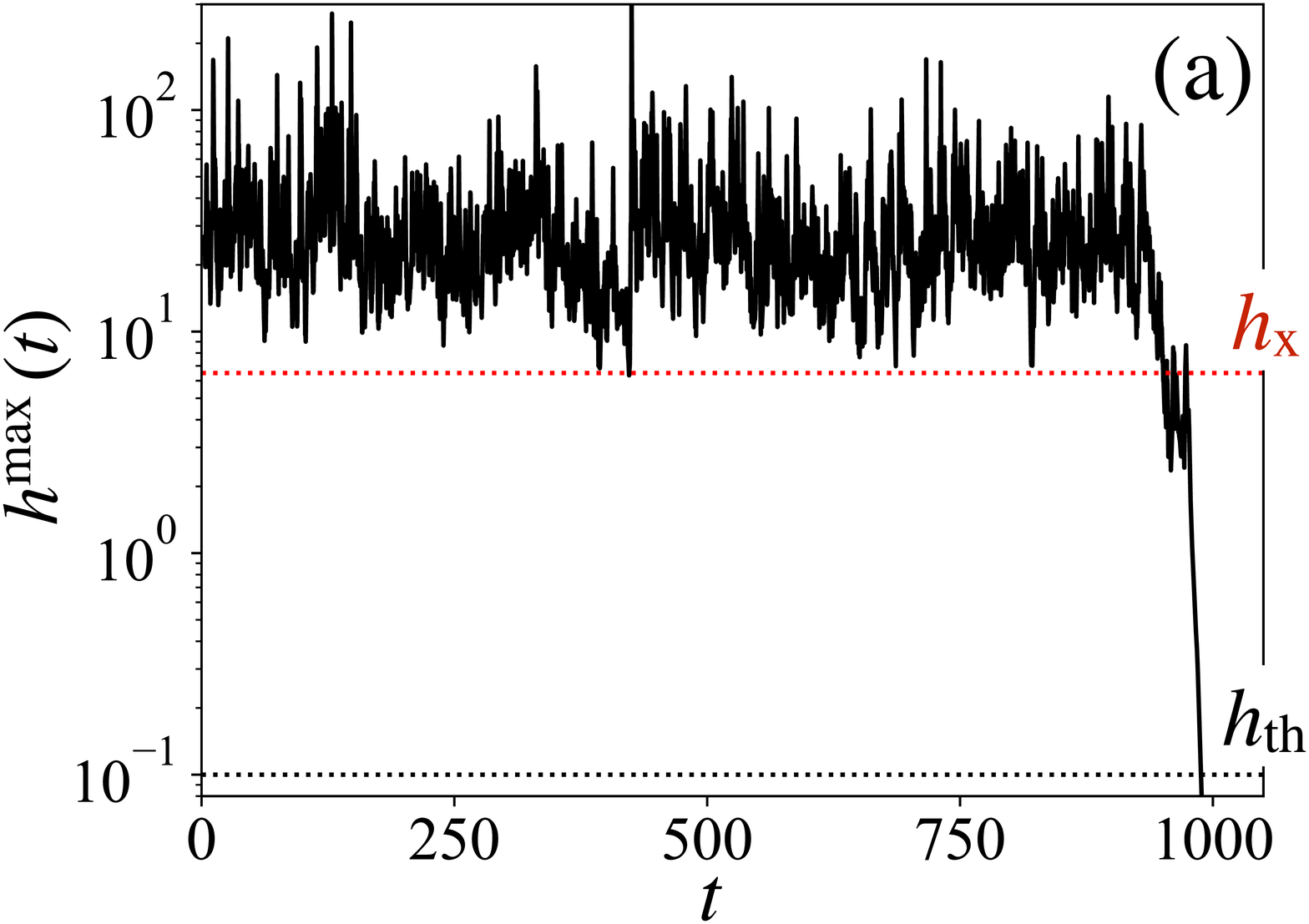}
\includegraphics[width=67mm]{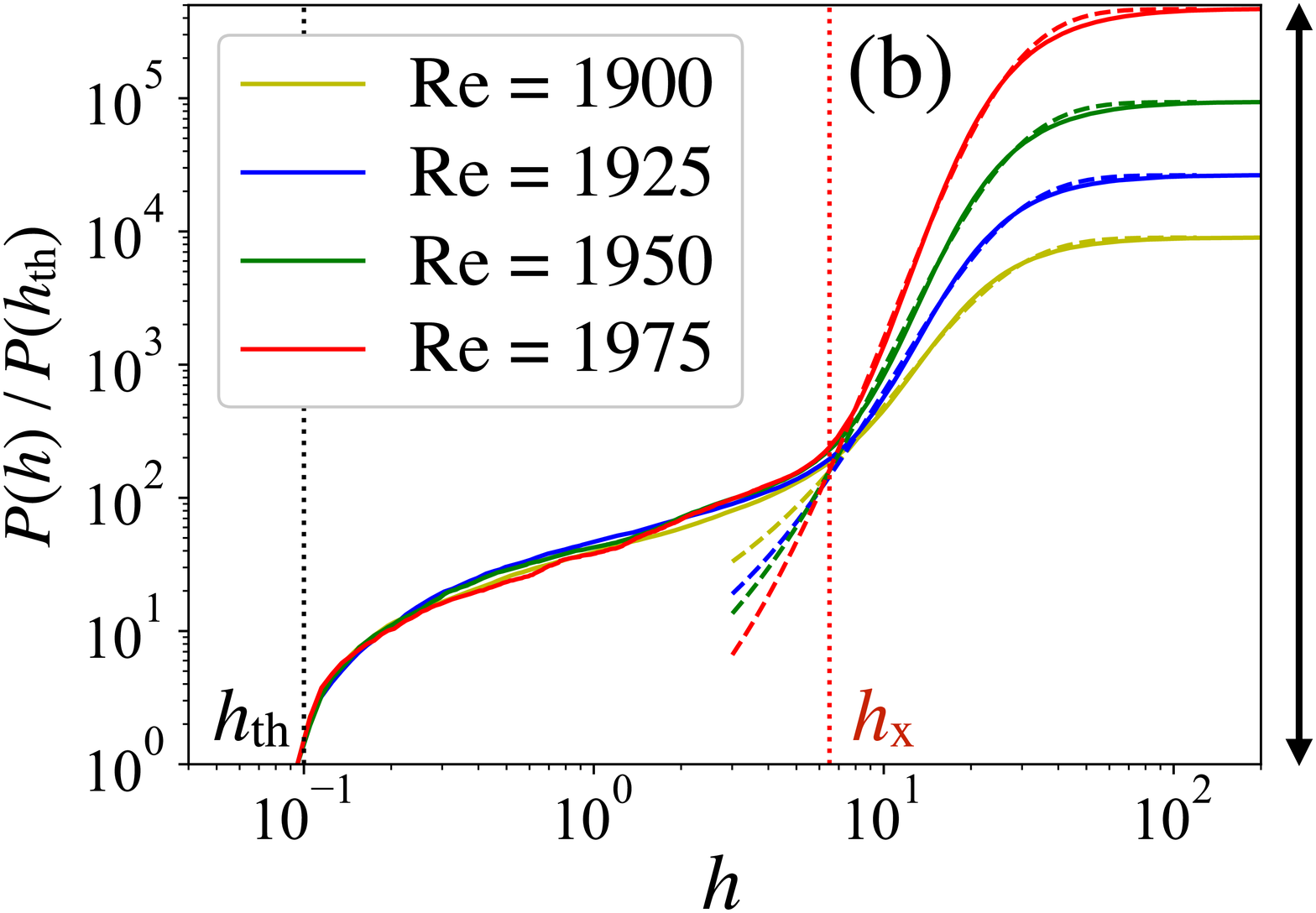}
\caption{\label{fig:cumumaximum_a}  
{\bf (a)} A trajectory of the maximum turbulence intensity $h^{\max}(t)$ for $L=50D$ and $\Reynul=1900$, where the puff decays when $t\simeq 1000$. We set $h_{\rm th}=0.1$, below which the maximum turbulence intensity always monotonically decreases ({\it i.e.}, puffs always decay). This threshold line is shown as a black dotted line. 
{\bf (b)} The CDF of the maximum turbulence intensity $P(h)$ (\ref{eq:cumulative}) divided by its minimum value $P(h_{\rm th})$, obtained from numerical simulations (solid coloured lines). The measurement interval $\delta t$ is set to 1/4. The pipe length is set to $L=50D$. (For $L=100D$, similar results are obtained). Below a certain value $h_{\rm x}$, the function $P(h)/P(h_{\rm th})$ becomes independent of Reynolds number. This $h_{\rm x}$ is shown as a vertical red dotted line, being around 6.5 for $L=50D$ and 7.5 for $L=100D$.  
For  $h>h_{\rm x}$, $P(h)$ is well approximated by the Gumbel function $P_{\rm Re}(h)$ (\ref{eq:gambel}): 
We fit $P_{\rm Re}(h)/P(h_{\rm th})$ to $P(h)/P(h_{\rm th})$ for $h>h_{\rm x}$ and show them as  coloured dashed lines. See Table~\ref{Table:fittingGumbel} for the fitting parameters $\gamma$ and $h_0$ used in this figure. Note that the value of $1/P(h_{\rm th})$ is represented as the length of the double-headed arrow  (for $\Reynul=1975$). From this figure, we estimate $\Pi(h_{\rm x}) \equiv P(h_{\rm x})/P(h_{\rm th}) \simeq 158.8$ for $L=50D$ and 219.2  for $L=100D$.
  }
\end{center} 
\end{figure}

To this goal, we characterise the intensity of turbulence by using the squared z-component of vorticity $H(\bv r,t)=(\nabla \times \bv u)_z^2$. (Note that there are  other quantities that can characterise the turbulence intensity, such as the kinetic energy fluctuations -- see Appendix~\ref{appB} for the result obtained using this latter quantity.)   
We then consider the maximum value of this turbulence intensity over the pipe (maximum turbulence intensity),
\begin{equation}
\label{eq:turbulenceintensitymax}
h^{\rm max}(t) = \max_{\bv r}H(\bv r,t).
\end{equation}
Once this quantity goes below a certain threshold $h_{\rm th}$, $h^{\rm max}(t)$ monotonically decreases, leading to a quick exponential decay of the puff ({\it i.e.}, the puff dynamics shows transient chaos \citep{PhysRevE.84.016309,PhysRevLett.101.214501}). See Fig.~\ref{fig:cumumaximum_a}(a) for a decaying trajectory of $h^{\rm max}(t)$. In order to measure the CDF of $h^{\rm max}(t)$ in our numerical simulations, we save the value of $h^{\rm max}(t)$ for every fixed time interval $\delta t$  
\footnote{We stop saving $h^{\rm max}(t)$ once  a value of $h^{\rm max}(t)$  satisfying $h^{\rm max}(t) < h_{\rm th}$ is saved.}.  We denote by $(h^{\max}(t_i))_{i=1}^{N}$ the data obtained from all the simulations with the number of saving points $N$. By using this data, we then define a CDF $P(h)$ as the probability that $h^{\rm max}(t)$ takes a value less than or equal to $h$:
\begin{equation}
P(h) =  \sum_{i=1}^{N}\frac{\Theta (h - h^{\max}(t_i))}{N}
\label{eq:cumulative}
\end{equation}
with a Heaviside step function $\Theta(h)$. 
Note that  the number of decay events $n_{\rm d}$ is equal to $NP(h_{\rm th})$ because 
$n_{\rm d} = \sum_{i=1}^{N} \Theta (h_{\rm th} - h^{\max}(t_i)) = NP(h_{\rm th})$. This indicates that the decay time $\tau_{\rm d}\equiv ~ (N\delta t)/n_{\rm d}$ is expressed as \citep{PhysRevE.97.022207}
\begin{equation}
\tau_{\rm d} = \frac{\delta t}{P(h_{\rm th})},
\label{eq:taudecay}
\end{equation}
when $N$ is sufficiently large. We set $h_{\rm th}=0.1$ throughout this article\footnote{Precisely, the threshold value  $h_{\rm th}$ should be defined as the value below which puffs always decay monotonically, but above which puffs have a certain probability to grow up and survive (even if this probability is very small). In reality, it is not easy to determine this precise value $h_{\rm th}^*$ from numerical simulations. Fortunately, the magnitude of the errors (due to an underestimation of $h_{\rm th}<h^*_{\rm th}$) does not depend on the Reynolds number and it is proportional to $\log(h^*_{\rm th}/h_{\rm th})$ because of the exponential decay of puffs when $h_{\rm th}$ is very small. When we consider $\rm Re$ close to its critical value, these errors are negligible as $\tau_{\rm d}$ increases super-exponentially. }.

We measure this CDF $P(h)$ in DNS, rescale it by the threshold probability $P(h_{\rm th})$, and plot it in Fig~\ref{fig:cumumaximum_a}(b). When $h$ is smaller than a certain value $h_{\rm x} \ (>h_{\rm th})$, the overlap of $P(h)/P(h_{\rm th})$
between different Reynolds numbers is observed, {\it i.e.}, $P(h)/P(h_{\rm th}) \simeq \Pi (h)$ for $h<h_{\rm x}$ with a Re-independent function $\Pi (h)$. (Note that $h_{\rm x} \simeq 6.5$ for $L=50D$ and $h_{\rm x} \simeq 7.5$ for $L=100D$ from Fig~\ref{fig:cumumaximum_a}(b)\footnote{From this tendency, we expect that $h_{\rm x}$ increases gradually as $L$ increases, which eventually converges to a certain value. This convergence can be studied in numerical simulations with the Reynolds number that is  relatively far away from the critical value (e.g., $\rm Re = 1900$) -- this is an interesting future study.}). When $h^{\rm max}(t)$ is smaller than $h_{\rm x}$, dynamics are in a metastable state where the puff is hovering between death and life (see the panel (a) of Fig. \ref{fig:cumumaximum_a}). This overlap indicates that  the dynamics in this metastable state are independent of (or less sensitive to) the change of the Reynolds number. 
On the other hand, when $h$ is greater than this value, we find that $P(h)$ is well-described by the Gumbel distribution function $P_{\Reynul}$
\begin{equation}
P_{\Reynul}(h) \equiv \exp\left [  - \exp\left ( - \gamma (h-h_0) \right )  \right ],
\label{eq:gambel}
\end{equation}
where $\gamma$, $h_0$ are fitting parameters depending on $\Reynul$.  In summary, we have confirmed that the scaled probability $P(h)/P(h_{\rm th})$ is well approximated as
\begin{equation}
\frac{P(h)}{P(h_{\rm th})} \simeq 
\begin{cases}
\Pi (h)   &  \quad  h \leq h_{\rm x}  \\
\ P_{\Reynul}(h)/ P(h_{\rm th})  &  \quad  h > h_{\rm x}.
\end{cases}
\label{eq:ph}
\end{equation}

\begin{figure}
\begin{center}
\includegraphics[width=67mm]{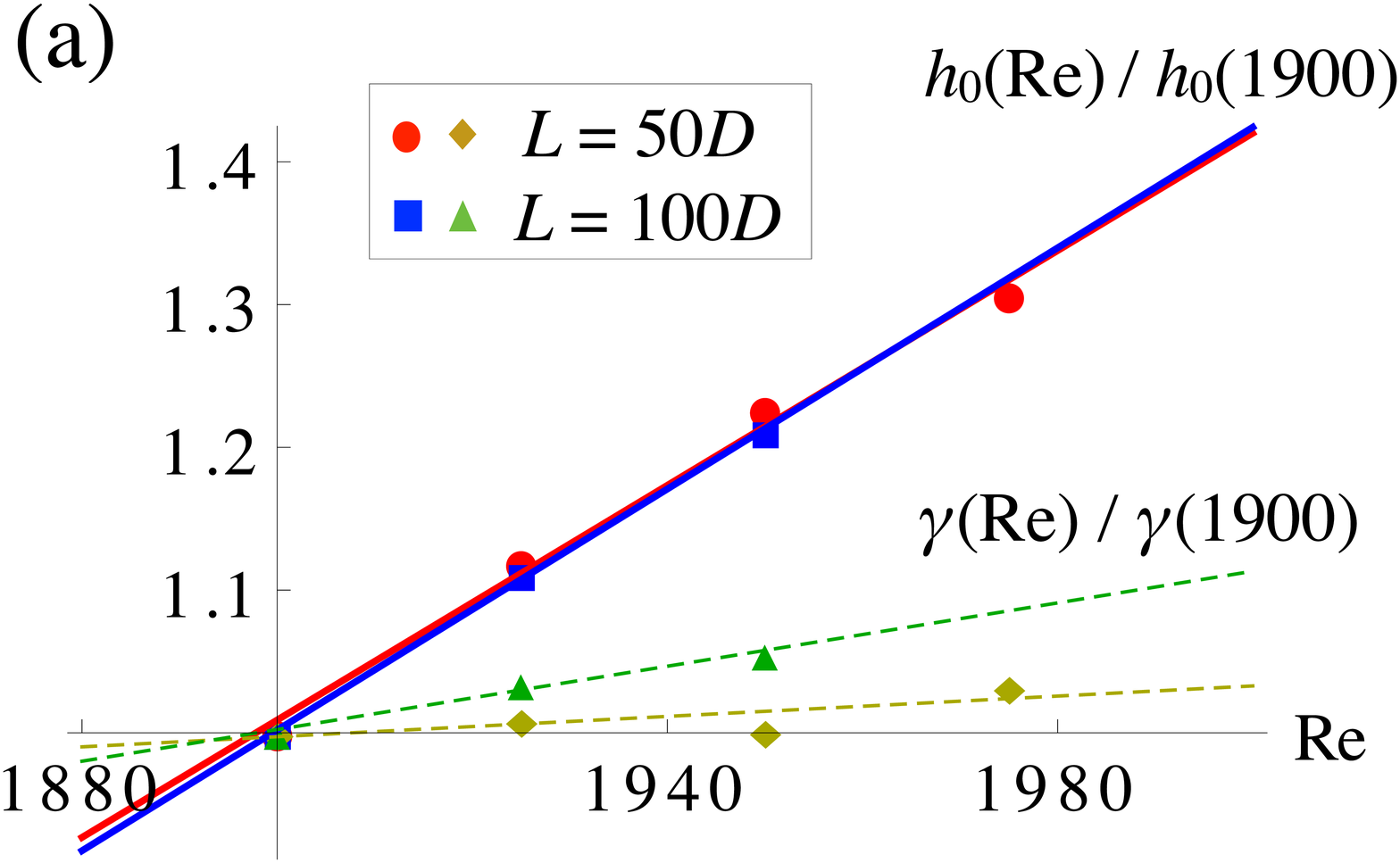}
\includegraphics[width=67mm]{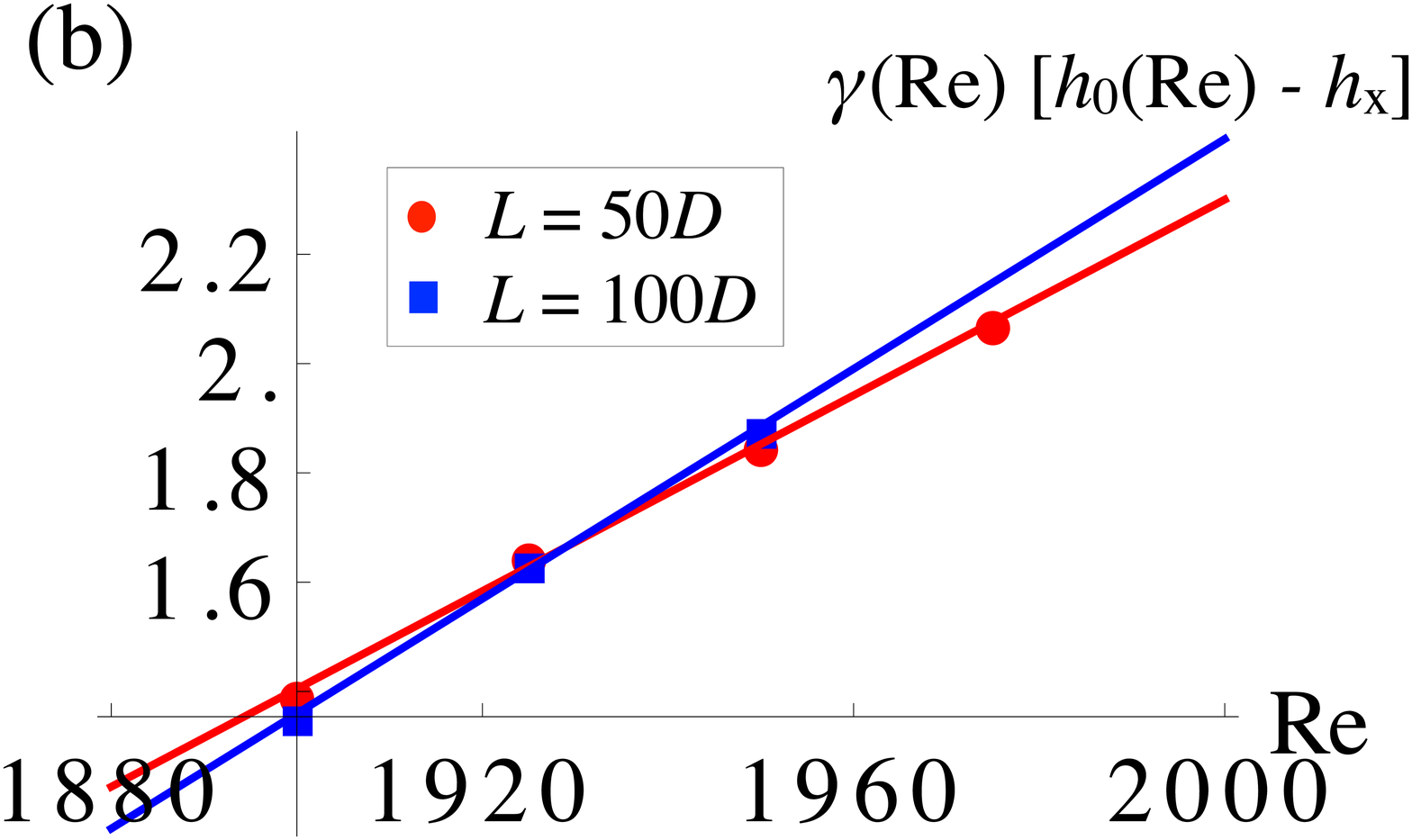}
\caption{\label{fig:cumumaximum_c}  
{\bf (a)} The fitting parameters $\gamma$ and $h_0$ in the Gumbel function (\ref{eq:gambel}), used in Fig.~\ref{fig:cumumaximum_a}(b), as a function of $\rm Re$. Red circles, yellow diamonds, blue squares and green triangles correspond to ($h_0$ with $L=50D$), ($\gamma$ with $L=50D$), ($h_0$ with $L=100D$) and ($\gamma$ with $L=100D$), respectively. To plot them together in the same panel, we divide each $\gamma({\rm Re})$ and $h_0({\rm Re})$ by $\gamma(1900)$ and $h_0(1900)$. 
They show linear dependence on Reynolds number: the solid and dashed straight lines are the linear fit. See Table~\ref{Table:fittingvalues} for the slopes and intercepts of these linear  lines.   Note that the slopes of the fitting lines for $h_0/h_0(1900)$ are much larger than those for $\gamma/\gamma(1900)$.
{\bf (b)} $ \gamma({\rm Re}) [h_0({\rm Re})-h_{\rm x}]$ as a function of ${\rm Re}$, where $ \gamma({\rm Re})$ and $h_0({\rm Re})$ use the same values in the panel (a), and $h_{\rm x}=6.5$ for $L=50D$ and $h_{\rm x}=7.5$ for $L=100D$. We find a linear dependence: $ \gamma({\rm Re}) [h_0({\rm Re})-h_{\rm x}] = a \ {\Reynul} + b$ (\ref{eq:linearh0hx}).  Here $a$ and $b$
are determined as ($a=0.00895442$, $b=-15.6087$) for $L=50D$ and ($a=0.0105166$, $b=-18.6222$) for $L=100D$. }
\end{center} 
\end{figure}

Based on this observation, the double-exponential increase of the mean decay time is justified as follows: 
From the continuity condition of (\ref{eq:ph}) at $h=h_{\rm x}$, we get  $\Pi(h_{\rm x}) = P_{\rm Re}(h_{\rm x})/ P(h_{\rm th})$.
Using the relation (\ref{eq:taudecay})  that connects  the decay time $\tau_{\rm d}$ and $P(h_{\rm th})$, we then derive
\begin{equation}
\tau_{\rm d} = \delta t / P(h_{\rm th}) = \delta t \ \Pi(h_{\rm x}) \ \exp\left [  \exp\left (  \gamma (h_0-h_{\rm x}) \right )  \right ].
\label{eq:doubleexponential_before}
\end{equation}
In the right-hand side of this expression, the $\Reynul$-dependence only comes from $\gamma$ and $h_0$, because $h_{\rm x}$ does not depend on $\rm Re$ (at least in the range of $\rm Re$ we consider). In order to study the $\rm Re$-dependence on $\gamma$ and $h_0$, we next plot these quantities as a function of $\Reynul$ in Fig.~\ref{fig:cumumaximum_c}(a). The figure indicates that, within the examined range, these parameters depend linearly on $\Reynul$.
Especially, we can see that the slope of the linear fitting curve for $\gamma$ is much smaller than that for $h_0$, which means we can approximate $\gamma h_0$ as a linear function of $\Reynul$. We thus get
\begin{equation}
 \gamma (h_0-h_{\rm x}) \simeq  a \ {\Reynul} + b
\label{eq:linearh0hx}
\end{equation}
with coefficients $a$ and $b$. In Fig.~\ref{fig:cumumaximum_c}(b), we confirm this linear approximation by plotting
the left-hand side of (\ref{eq:linearh0hx}) as a function of $\rm Re$. We also determine the coefficients $a$ and $b$ from this figure, and summarise them in the caption of Fig.~\ref{fig:cumumaximum_c}(b). 
Finally, using (\ref{eq:linearh0hx}) in (\ref{eq:doubleexponential_before}), we derive 
the double-exponential formula 
\begin{equation}
\tau_{\rm d} \simeq \delta t  \ \Pi(h_{\rm x}) \exp\left [  \exp\left ( a \ {\Reynul} + b \right )  \right ].
\label{eq:tau_decay_last}
\end{equation}
Using $\delta t = 1/4$ and the values of $\Pi(h_{\rm x})$ measured in Fig.~\ref{fig:cumumaximum_a}(b) (together with $a,b$ obtained in Fig.~\ref{fig:cumumaximum_c}(b)), we plot this double exponential formula in Fig.~\ref{fig:decaytime} as red and blue solid lines. These theoretical lines are consistent with the direct measurements (red circles and blue squares). Note that the parameters $\gamma, h_0$ are determined from the measurements up to $\Reynul=1975$ for $L=50D$ and up to $\Reynul=1950$ for $L=100D$. But the obtained curves agree with the direct measurements for $\Reynul=2000$ with $L=50D$ and  for $\Reynul=1975$ with $L=100D$. This observation indicates that our method can be used to infer statistical properties in higher Reynolds numbers from the data obtained in lower Reynolds numbers.

\subsection{Relevance with the extreme value theory}
\label{Relevance_theory}

Our formula \eqref{eq:tau_decay_last} is the product of (i) the double exponential term $\exp [ \exp ( a  {\rm Re} + b )]$ and (ii) the constant term (independent of Re) $\Pi(h_x) \delta t$. The first double-exponential term is attributed to the probability of the event in which $h_{\max}$ of a fully-developed puff is weakened to $h_{\rm x}$. After this event, the weakened puff undergoes meta-stable dynamics as shown in the time series in Fig.\ref{fig:cumumaximum_a} (a). These meta-stable dynamics are known as edge states \citep{PhysRevLett.108.214502}. On the other hand, the second term corresponds to the average time for these meta-stable puffs to completely decay. In the original conjecture in \citep{PhysRevE.81.035304,Goldenfeld2017}, this second term was not considered. However, when the Reynolds number is close to its critical value, we expect that the double exponential term becomes dominant, so that their argument is still reasonable.

As seen in the previous subsection, the first term (i) was derived from the Gumbel distribution function introduced in the approximation \eqref{eq:ph}. The validity of this Gumbel distribution is where the extreme value theory could be relevant: The Fisher-Tippett-Gnedenko (FTG) theorem ensures that the CDF of the maximum value of a set of independent stochastic variables becomes either the Gumbel distribution function (\ref{eq:gambel}), Fr\'echet distribution function, or Weibull distribution function \citep{fisher1928limiting,gumbel1935valeurs}. Inspired by this theorem, Goldenfeld {\it et al} conjectured in \citep{PhysRevE.81.035304,Goldenfeld2017} that the Gumbel distribution function described the maximum turbulence intensity. In this work,  we confirmed that this conjecture was approximately satisfied.

\subsection{Close-up of the Gumbel-distribution approximation}
\label{Relevance_extreme_value_distributions}

The approximation of the CDF using the Gumbel distribution \eqref{eq:ph} was the key to the derivation of the double exponential formula  \eqref{eq:tau_decay_last}. Interestingly, in a detailed look at Fig.~\ref{fig:cumumaximum_a}(b), we detect small errors in this approximation (for $h$ larger than 20), though the errors are too small to affect the derivation of \eqref{eq:tau_decay_last}.  In general, approximations are better in CDFs than in probability distribution functions (PDFs) defined as the derivative of CDFs, because taking the derivative magnifies the errors. In this subsection, we  thus investigate  the corresponding PDF in order to understand the origin of  these errors. This is important for future studies to apply the same formulation to different problems or to investigate the same problem with a higher Reynolds number.

\begin{figure}
\begin{center}
\includegraphics[width=134mm]{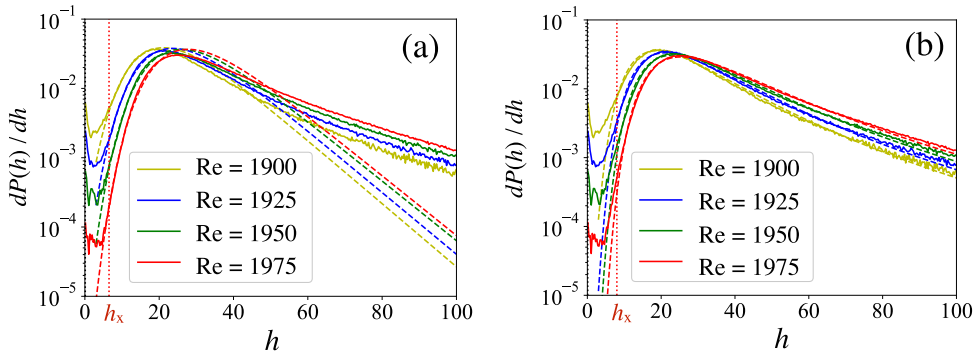}
\caption{\label{Fig:Prob:maxim}  
Probability distribution function of the maximum turbulence intensity $dP(h)/dh$ (where $P(h)$ is the CDF defined in (\ref{eq:cumulative})). The Gumbel distribution function \eqref{eq:gambel} and the Fr\'echet distribution function \eqref{eq:Frechet} are fitted to these curves and plotted, respectively, in the panel (a) and in the panel (b) as dashed lines.}
\end{center} 
\end{figure}

We  study the PDF of the maximum turbulence intensity $dP(h)/dh$, by plotting it together with the corresponding Gumbel probability distribution $dP_{\rm Re}(h)/dh$ in Fig.~\ref{Fig:Prob:maxim}(a). We can detect clear discrepancies between the PDF and the Gumbel distribution function for $h>20$, to which we attribute two reasons. First, the condition of the FTG theorem may not be exactly satisfied. For example, the turbulence-intensify field is not spatially independent,  whereas the independence is required for the FTG theorem to be applied\footnote{The original FTG theorem requires the variables to be independent but this condition can be weakened \citep{charras2013extreme}. It is anyway difficult to justify it with our turbulence intensify {\it a priori}.}. Second, assuming FTG theorem is applicable, the Gumbel distribution function may not be the correct one: it may be either the Fr\'echet distribution function or the Weibull distribution function describing the maximum turbulence intensify fluctuations. We can easily rule out the Weibull distribution, as it is for random variables with an upper bound (which is not true in our turbulence-intensity field). The Fr\'echet distribution is defined as
\begin{equation}
P_{\Reynul}^{\rm F}(h) \equiv \exp\left [  - \kappa \  (h-\eta) ^{-\zeta} \right ],
\label{eq:Frechet}
\end{equation}
where $\kappa, \eta, \zeta$ are parameters. A characteristic of the Fr\'echet distribution is the power law decay of the logarithm of the CDF ({\it c.f.}, the exponential decay in the case of the Gumbel distribution). 
The FTG theorem states that, when the distribution function of each variable (the turbulence intensity field for each position in our problem) decays  in power laws, the maximum among these variables is described by the Fr\'echet distribution with $\zeta$ equal to the decay exponent of each variable ({\it c.f.}, when the distribution function of each variable decays exponentially, the maximum among these variables is described by the Gumbel distribution). We fit the Fr\'echet distribution to our PDF of the maximum turbulence intensity $dP(h)/dh$ by tuning the three parameters $\kappa, \eta, \zeta$. Surprisingly, the agreement in the Fr\'echet distribution is far better than the Gumbel distribution function, as shown in Fig.~\ref{Fig:Prob:maxim}(b). The PDF, $dP(h)/dh$, clearly shows a presence of the power law decay  for large turbulence intensity that can be explained by the Fr\'echet distribution, but not by the Gumbel distribution\footnote{
The PDF of the Gumbel distribution is $\gamma \exp [ - \gamma(h-h_0) - e^{-\gamma(h-h_0)} ]$ that decays exponentially when $h$ is large, while the PDF of the Fr\'echet distribution is $ \kappa \zeta (h-\eta) ^{-1-\zeta}  \exp\left [  - \kappa \  (h-\eta) ^{-\zeta} \right ]$ that decays in power law when $h$ is large. In Fig.~\ref{Fig:Prob:maxim}, we can clearly see that $dP(h)/dh$ decays slower than exponential (and decays in a power law).
}.

Nevertheless, as long as we consider the decay time, the errors in the Gumbel-distribution approximation in PDF are not important, because the CDF, which plays a key role in the derivation of the decay time\footnote{For example, as seen from \eqref{eq:taudecay}, $\tau_{\rm d}$ is directly related to $1/P(h_{\rm th})$ (as indicated as the double-headed arrow in Fig~\ref{fig:cumumaximum_a}(b)), which is hardly affected by tiny deviations between the Gumbel distribution and the CDF.}, is hardly affected by these errors. Indeed, replacing the Gumbel distribution function by the Fr\'echet distribution function in the argument of Section~\ref{sec:maximum_turbulence}  still leads to the same result  \eqref{eq:tau_decay_last} (see Appendix~\ref{What_is_different_with_Frechetdistribution}). But in the future, when we study the same problem with higher precisions (more statistics), or when we study the problem in higher Reynolds numbers, we could possibly detect the deviations from the double-exponential formula  \eqref{eq:tau_decay_last}. In that case, using the Fr\'echet distribution function for more accurate analysis will be interesting.

\section{Conclusion} 
In this paper, using a large number of DNS, we measure the decay time of turbulent puff in pipe flows up to $\Reynul=2000$. In DNS, periodic boundary conditions are employed in the axial direction, so that the insufficient length of the pipe introduces confinement effects on puff dynamics. Our numerical simulations show that, as the length of the pipe increases ({\it i.e.}, as the confinement effects disappear), the obtained decay times increase, resulting in values larger than those used in \citep{Avila192} for large Reynolds numbers (that were not previously studied using DNS). We also note that the curvature of the double exponential curves between our theoretical line \eqref{eq:doubleexponential_before} and the one used in \citep{Avila192} is slightly different. 
Indeed in the inset of  Fig.~\ref{fig:decaytime}, $\log\tau_{\rm d}$ shows a straight line in the form used by \citep{Avila192}, while our theoretical expression \eqref{eq:tau_decay_last} shows a lightly curved line. The difference between the two formulas is a constant term $c$ by which the double exponential form is multiplied: $c \exp [ \exp [a {\rm Re} + b]]$ (with parameters $a,b$) where $c = 1$ for the form used in \citep{Avila192}, while $c = \delta t \Pi(h_{\rm x})$ for our expression\footnote{$\log\{ \log \{ c \exp \{ \exp [a {\rm Re} + b] \}  \} \} = \log [\log c +\exp [a {\rm Re} + b] ]$, which becomes a linear function of  ${\rm Re}$ only when $c=1$.}. Note that the value of the constant term changes in a different time unit, which indicates that $c=1$ in the convective time unit ($D/U_{\rm b}$) used in \citep{Avila192} is an approximation.

It was conjectured in \citep{PhysRevE.81.035304,Goldenfeld2017} that the double-exponential formula of the mean decay time could be derived using the Gumbel distribution function of maximum kinetic energy fluctuations based on the extreme value theory. We measure the CDF of the maximum turbulence intensity \eqref{eq:turbulenceintensitymax} and show that the function is indeed approximately described by the Gumbel distribution function \eqref{eq:gambel}, demonstrating that their conjecture is reasonable, in the range of the Reynolds number between 1900 and 2000. In Section~\ref{Relevance_extreme_value_distributions}, we also study the corresponding probability distribution function, and find that another extreme value distribution, the Fr\'echet distribution \eqref{eq:Frechet}, fits better than the Gumbel distribution, indicating the presence of  the power law decay in the distribution of the turbulence intensity field. This finding implies that the double-exponential formula of the decay time could be merely an approximation. It will be interesting to study the same problem with more detailed statistics or with a higher Reynolds number, as the true expression of the mean decay time could be possibly discovered using the same approach developed in this article.

Our approach will be also important for the future study in different problems of turbulence, where the extreme events trigger transitions. For example, a similar turbulence-decay problem has been studied in a different geometry (channel flows) in \citep{Shimizu_2019}, where they observed the agreement between the probability distribution of a turbulence intensity and the Gumbel distribution in a linear-scale plot. It would be interesting to ask if the Fr\'echet distribution fits better in their case as well with more detailed statistics. Furthermore,  turbulent transitions between 2D- and 3D-dynamics have been long studied \citep{smith1996crossover, celani2010morethantwo, benavides_alexakis_2017, musacchio2017split,alexakis2018cascades}, where 
a super-exponential increase of the transition time was  observed in thin-layer turbulent condensates \citep{van_kan_nemoto_alexakis_2019}. This super-exponential increase could be also studied using the extreme value theory.

Studying these super exponential problems in DNS is computationally demanding. A brute-force approach using a large number of DNS is efficient as proven in this work.  But exploiting so-called rare-event sampling methods could be also helpful. Such sampling methods include instanton methods based on Freidlin-Wentzell theory \citep{Chernykh2001, Heymann2008, Grafke_2015_2, Grafke_2015, Grigorio_2017} as well as splitting methods that simulate several copies in parallel \citep{Allen_2005, Giadina_2006, cerou2007adaptive, tailleur2007probing, Teo_2016, Nemoto_2016, Lestang_2018, bouchet2018rare}. These methods have been successfully applied to many high-dimensional chaotic dynamics and proven invaluable.

\section*{Acknowledgements}
The authors thank Dwight Barkley and Laurette Tuckerman for fruitful discussions and comments. This work was granted access to the HPC resources of CINES/TGCC under the allocation 2018-A0042A10457 made by GENCI (totally 3 million hours) and of MesoPSL financed by the Region Ile de France and the project Equip@Meso (reference ANR-10-EQPX-29-01) of the program Investissements d'Avenir supervised by the Agence Nationale pour la Recherche.

\section*{Declaration of interests}
The authors report no conflict of interest.

\appendix
\section{Simulation detail}\label{appA}
\begin{figure}
\begin{center}
\includegraphics[width=70mm]{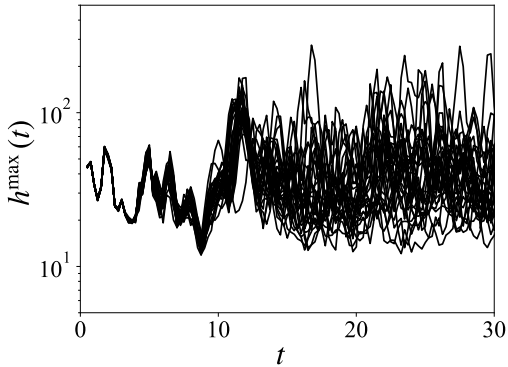}
\caption{\label{fig:divergence_trajectories}  
Plots of 30 different $h^{\max}$ trajectories  starting from different initial conditions for ${\rm Re} = 1900$ and $L=50D$. We add a  small Gaussian noise to a configuration containing a single puff, to generate different initial conditions. After the Lyapunov time, the effect of the small noise becomes exponentially important and make the dynamics with slightly different initial conditions be independent. Here we see that $t=20 \sim 30$ is large enough to make the dynamics independent. For  the Reynolds numbers higher than 1900, we expect that the Lyapunov time is much smaller, as the dynamics is more chaotic. 
 }
\end{center} 
\end{figure}

\begin{figure}
\begin{center}
\includegraphics[width=70mm]{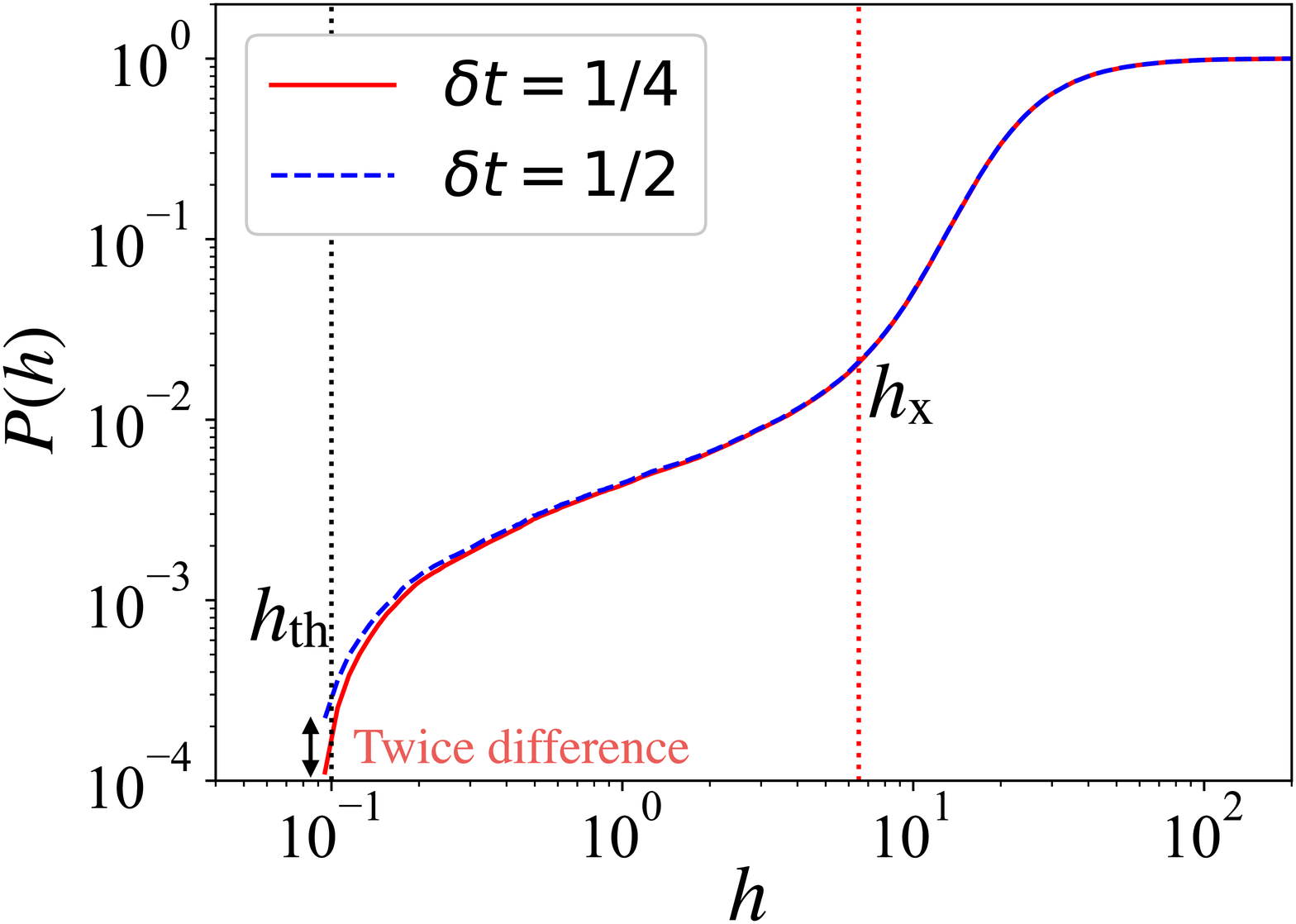}
\caption{\label{fig:different_deltat}  The CDF $P(h)$ for different saving times $\delta t $ for ${\rm Re}=1900$, $L=50D$. We observe that changing $\delta t $ only affects the functional shape of $P(h)$ close to $h_{\rm th}$ without changing $P(h)$ for $h\geq h_{\rm x}$. In this figure, $P(h_{\rm th})|_{\delta t =1/4} = 0.0001113$ and $P(h_{\rm th})|_{\delta t =1/2} = 0.0002226$, so we get $2 P(h_{\rm th})|_{\delta t =1/4} =  P(h_{\rm th})|_{\delta t =1/2} $. This relation is generally true as explained in the last bullet point of Appendix \ref{appA}: Multiplying $\delta t $ by an arbitrary factor results in multiplying $P(h_{\rm th})$ by the same factor. This means that the average decay time $\tau_{\rm d} (= \delta t / P(h_{\rm th}))$ \eqref{eq:taudecay} is independent of the value of $\delta t $. 
 }
\end{center} 
\end{figure}

We used an open source code openpipeflow \citep{WILLIS2017124}, which simulates
flows in a cylindrical domain by solving the Navier-Stokes equations. 
Below are the summary of the parameters and the settings we used:
\begin{itemize}
\item 
For azimuthal and longitudinal directions of the pipe, 
the spectral decomposition is used to evaluate the derivatives, for which we use 24 variables for the azimuthal direction and 384 variables (for $L=50D$) or 768 variables (for $L=100D$) for the longitudinal directions.
For the radial direction, a finite-element method is used, for which the radial space is divided into 64 points using Chebyshev polynomials. 

\item The code can solve the Navier-Stokes equations under two conditions: fixed flux conditions and fixed pressure conditions. In the present study we used the fixed flux condition for the simulations. 

\item For the time step, the algorithm uses a second-order predictor-corrector
scheme with automatic time-step control with courant number 0.5.

\item We define that a puff decays when $h^{\rm max}<0.1$ is satisfied. We never observed that a puff regenerates once $h^{\rm max}<0.1$ is observed.

\item To prepare initial velocity fields, we use a configuration where a single steady puff already exists. We added a small Gaussian noise to this configuration, and simulate it during a time interval 50 or 100. These time intervals are our initial relaxation time -- see Fig.~\ref{fig:divergence_trajectories} for how the trajectories of $h^{\max}$ with different initial conditions deviate with each other. From the figure, we confirm that $t=20\sim 30$ for ${\rm Re} = 1900$ and less for ${\rm Re} > 1900$ could be large enough to make the dynamics independent, so that our initial relaxation interval (50 or 100) is reasonable. We also checked that both initial relaxation intervals (50 and 100) show consistent (almost the same) results for ${\rm Re}\geq 1900$ \footnote{When an initial relaxation interval is sufficiently larger, the time at which we observe the decay of puff (measured from the initial relaxation time) is distributed as the memoryless exponential probability. This memoryless distribution leads to  the Poisson distribution for the number of decays, as we used in the main text \eqref{eq:poissondist}. Increasing further the initial relaxation time simply results in reducing the amount of data we gathered. }. 

\item We set the time interval of saving data $\delta t$ to be 1/4 (in units of $D/U_{\rm b}$) for the simulations in Fig.~\ref{fig:cumumaximum_a}(b). Note that $P(h_{\rm th}) = 1/N$ (from \eqref{eq:cumulative}), where $N$ is the number of saving points. Since $N$ is inversely proportional with the saving interval $\delta t$, we find that multiplying $\delta t$ by an arbitrary factor $c$ simply changes the probability $P(h_{\rm th})$ to $P(h_{\rm th}) c$. The decay time $\tau_{\rm d}$ is then independent of $\delta t$, because $\tau_{\rm d} = \delta t / P(h_{\rm th})$ \eqref{eq:taudecay}. 
Furthermore, in our theoretical expression of $\tau_{\rm d}$  in  \eqref{eq:doubleexponential_before}, the change of $\delta t$ is compensated by the change of $\Pi(h_{\rm x})$ without affecting the double exponential term. This can be seen from a numerical example of $P(h)$ in Fig.~\ref{fig:different_deltat} -- changing $\delta t$ only alters the functional shape of $P(h)$ close to $h_{\rm th}$, which leaves $P(h)$ for $h\geq h_{\rm x}$ unchanged. Multiplying $\delta t$ by an arbitrary factor $c$ thus changes $\Pi(h_{\rm x})(\equiv P(h_{\rm x})/P(h_{\rm th}))$ to $\Pi(h_{\rm x}) /c$, which compensates the change of $\delta t$ in  \eqref{eq:doubleexponential_before}. 

\end{itemize}

\subsection{Parallelisation}\label{appA-1}

Although the open source code openpipeflow has an option to use multiple threads for a single pipe flow simulation, relying on it when we use a considerably large number of threads would not be efficient. What we are interested in are the statistical properties of the decay events of the puff, which can be accumulated in independent simulations. Instead of parallelising a single pipe flow simulation, we thus simulate many pipe flows in an {\it embarrassingly parallel} way, where each thread is used to simulate a single pipe flow. Here we explain the detail of the architecture of this parallelisation.

\begin{figure}
\begin{center}
\includegraphics[width=120mm]{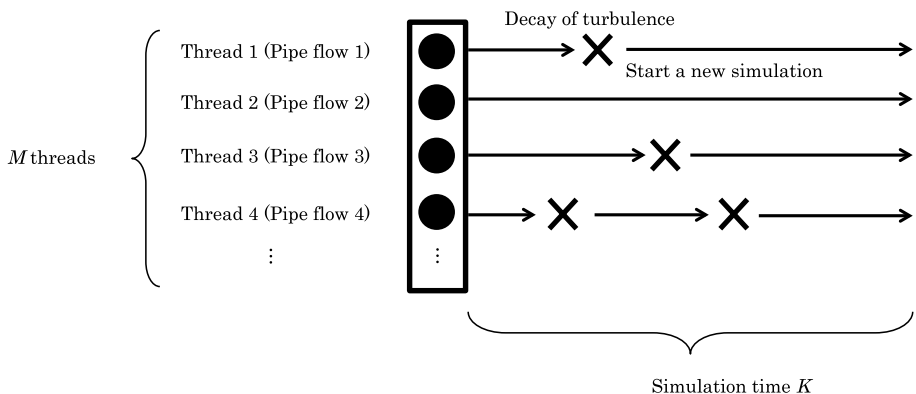}
\caption{\label{fig:structure_2}  
We simultaneously simulate $M$ pipe flow simulations with different initial conditions in $M$ threads, where each pipe flow simulation is allocated to a single thread.  When a decay of puff is observed in a thread, we immediately launch another pipe flow simulation in that thread with a different initial condition. The simulations continue during a time interval $K$. The values of these $M$ and $K$ are summarised in Table \ref{Table:M_and_K}. }
\end{center}
\end{figure}

As shown in Fig.~\ref{fig:structure_2}, we perform $M$ independent pipe flow simulations. Each pipe flow simulation is allocated to a single thread, thus $M$ threads are required for these simulations. The initial conditions of these simulations are slightly different from each other, so that the simulations become independent after an initial relaxation interval. During these simulations, when a puff decays in a thread, we immediately launch another pipe flow simulation in that thread using a new initial condition (that contains a developed puff but different from the other initial conditions that have been already used). All threads are thus always occupied. We continue these simulations for a computational time interval $K$ (or $K_{\rm conv}$ in convective time units). We summarise in Table~\ref{Table:M_and_K} the values of $M$ and $K$ we use, as well as the number of decay events $n_{\rm d}$ we observe.

\begin{table}
\begin{center}
          \caption{\label{Table:M_and_K} The number of threads $M$, the simulation time interval for each thread ($K$ in wall time and $K_{\rm conv}$ in convective units $D/U_{\rm b}$), the number of decaying events $n_{\rm d}$ (for all threads during the simulation interval $K$) and the total simulation interval $T$ in convective units (for all  threads during the simulation time $K$, excluding the initial relaxation time). $K_{\rm conv}$ slightly fluctuates depending on the thread, so the average value over all the threads is used ({\it i.e., } $K_{\rm conv} \equiv T / M$). Note also that $K_{\rm conv}$  do not include the initial relaxation time, while $K$ does. For the simulations, we used two cluster machines: \href{https://www.cines.fr/calcul/materiels/occigen/}{CINES OCCIGEN} and \href{http://www-hpc.cea.fr/fr/complexe/tgcc-JoliotCurie.htm}{TGCC Joliot Curie KNL Irene}, where $({\rm Re}, L) = (1900, 50), (1925, 50), (1950, 50), (1900, 100), (1925, 100), (1950, 100), (2040, 50)$ used CINES-OCCIGEN, while $({\rm Re}, L) = (1975, 50), (2000, 50)$ used TGCC-IreneKNL. $({\rm Re}, L) = (1975, 100)$ used the combined data obtained from both machines.
          }
          \begin{tabular}{cccccc} \hline
                        & $M $   & $K $ &  $K_{\rm conv}$ & $n_{\rm d}$  & $T$ 
            \\    \hline 
            $(L=50D)$ \vspace{0.1cm} \\ 
  ${\rm Re} = 1900 $    &       96                 & 13 days        &   1780.8     &  76  &  170958
            \\  
  ${\rm Re} = 1925 $   &       192                      &  13 days           &  1825.7     & 53 &  350528
              \\ 
   ${\rm Re} = 1950 $   &       384                      &  10 days           &  1404.8    &  23 & 539445
              \\
 ${\rm Re} = 1975 $   &       508                   &  47.75 days         &  2766.3    &     12 & 1405280
              \\ 
 ${\rm Re} = 2000 $   &       762                      &  27 days           &   1471.0   &      1  &  1120879
              \\
 ${\rm Re} = 2040 $   &       576                      &  18.7  days   &  $\sim 2400$  & 0   &   1382400               
              \\               \hline
            $(L=100D)$ \vspace{0.1cm} \\ 
  ${\rm Re} = 1900 $    &       192                  & 10 days         &  522.1    &   39 &   100241
            \\  
  ${\rm Re} = 1925 $   &       192                     &  36 days       &  2087.0    &     44   &  400705
              \\ 
   ${\rm Re} = 1950 $   &       384                      &  32 days      &  1869.2    &     18    &   717760
              \\
 ${\rm Re} = 1975 $   &       1530                      &  22.3 days     &   1005.5   &       4    &   1538470
              \\    \hline
          \end{tabular}  
        \end{center}
 \end{table}

\section{CDF of the maximum kinetic energy fluctuations}\label{appB}

\begin{figure}
\begin{center}
\includegraphics[width=70mm]{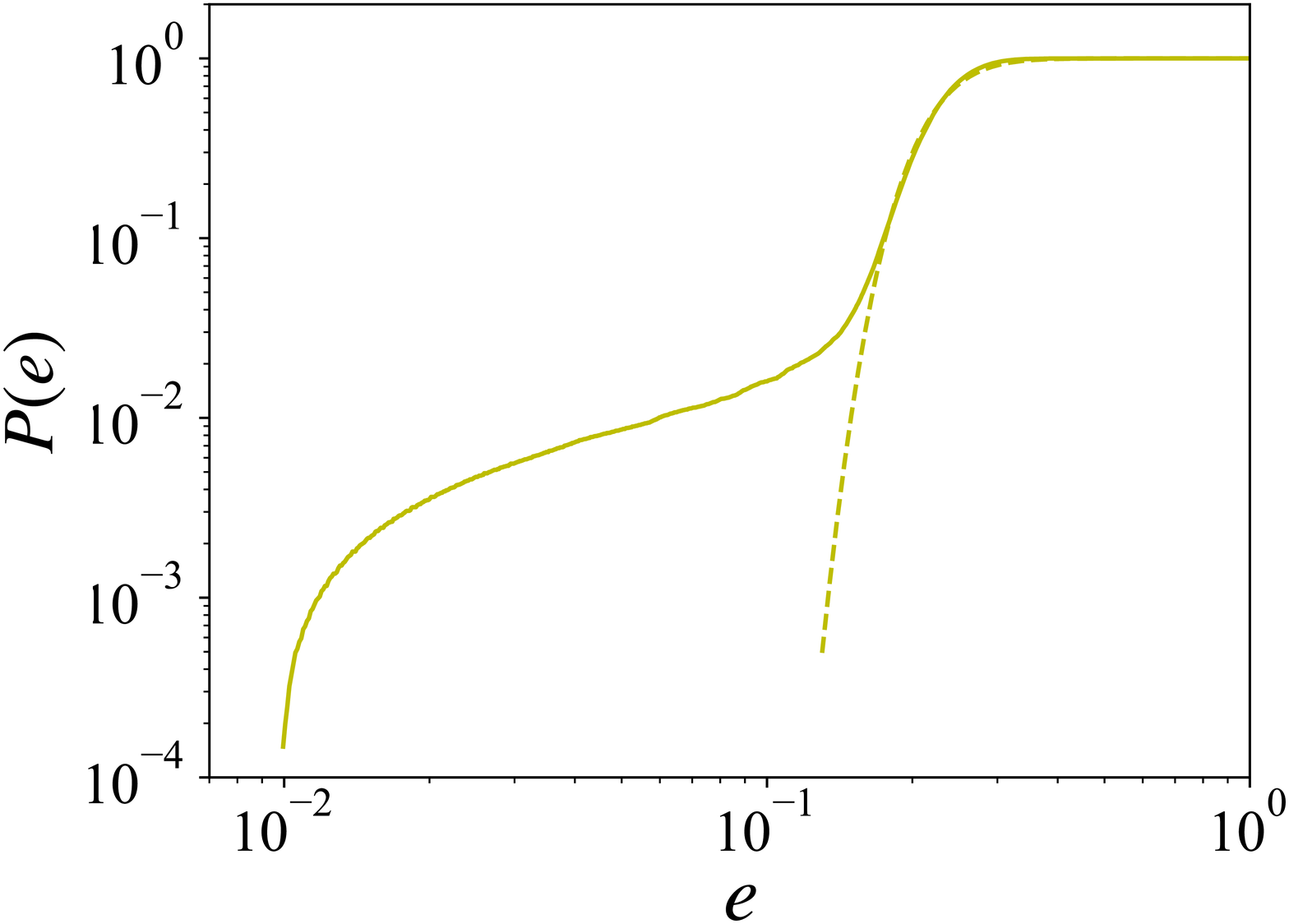}
\caption{\label{Fig:Prob_velocityfluctuations}  
The CDF of the maximum kinetic energy fluctuations $e^{\rm max}(t)$ defined in (\ref{eq:def:emax}), obtained from numerical simulations (solid line) with ${\rm Re}=1900$, $L=50D$.  The Gumbel distribution function (\ref{eq:gambel}) is fitted to this data and shown as a yellow dashed line. The parameters in the fitting curve are determined as 
$\gamma = 26.27,  h_0 = 0.2073$. 
}
\end{center} 
\end{figure}

We define the kinetic energy fluctuations $E$ as the kinetic energy of the flow fields without the contribution of the laminar flow $\bv u_{\rm lam}$: $E(\bv r, t)=\left [ \bv u (\bv r, t) - \bv u_{\rm lam} \right ]^2$. Similarly to $h^{\rm max}(t)$, we define the maximum  value of this kinetic energy fluctuations over the pipe as
\begin{equation}
e^{\rm max}(t) = \max_{\bv r} E(\bv r, t). 
\label{eq:def:emax}
\end{equation}
The CDF of this $e^{\max}(t)$ is defined in the same way as in \eqref{eq:cumulative}, denoted by $P(e)$. We measure $P(e)$ in numerical simulations and show it in Fig.~\ref{Fig:Prob_velocityfluctuations}. It shows qualitatively similar structures as those in Fig.~\ref{fig:cumumaximum_c}(b), namely the function is composed of the two parts: (i) the part that can be approximately described by the Gumbel distribution and (ii) the relatively flat part that is related to the dynamics of a metastable puff.

\section{Using the Fr\'echet distribution function still leads to \eqref{eq:tau_decay_last} }
\label{What_is_different_with_Frechetdistribution}

The approximation errors from the Gumbel distribution in the PDF do not affect the double exponential formula \eqref{eq:tau_decay_last}, as long as the errors are well-hidden in the corresponding CDF. This indicates that, even if we use the Fr\'echet distribution  to approximate $P(h)$, this will still leads to the same double exponential formula \eqref{eq:tau_decay_last}, at least as an approximation. We demonstrate it in this Appendix.

\begin{figure}
\begin{center}
\includegraphics[width=134mm]{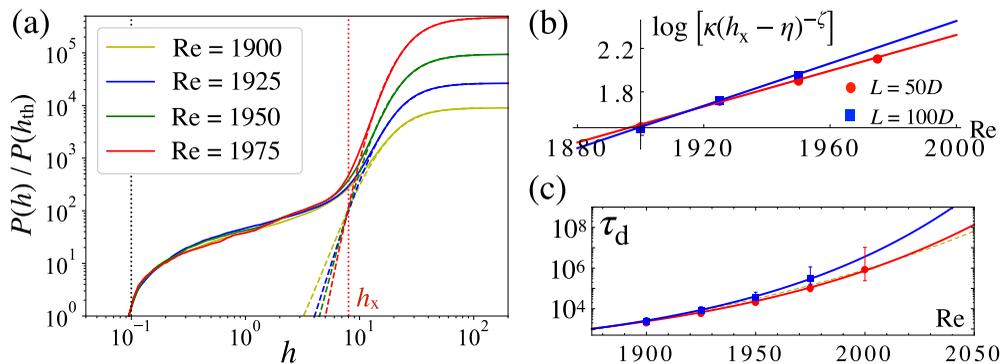}
\caption{\label{Fig:Frechet_1}  
{\bf (a)} The CDF of the maximum turbulence intensity $P(h)/P(h_{\rm th})$ in solid lines together with the fitted Fr\'echet distribution function in dashed lines ({\it c.f.} the same CDF with the fitted Gumbel distribution function in Fig.~\ref{fig:cumumaximum_a}(a)). The pipe length is $L=50D$ in this figure and a similar graph is obtained with $L=100D$. The values of $h_{\rm x}$, $\Pi (h_{\rm x}) \left (\equiv P_{\Reynul}^{\rm F}(h_{\rm x})/ P(h_{\rm th}) \right )$ are ($8, 108.2$) for $L=50D$ and ($8.5, 130.8$) for $L=100D$, respectively. {\bf (b)} $\log \left [\kappa (h_{\rm x} - \eta)^{-\zeta} \right ]$ as a function of Re, where $\kappa, \eta, \zeta$  are the fitting parameters of  the Fr\'echet distribution function \eqref{eq:Frechet} determined in the panel (a) for each Re. A straight line  \eqref{eq:logkappa_linear} is fitted to each plot and shown as a red solid line ($L=50D$) and a blue solid line ($L=100D$).  The fitting parameters $(\tilde a,\tilde b)$ are determined as $(-14.1817, 0.00825304)$ for $L=50D$ and $(-17.0837, 0.00976833)$ for $L=100D$. {\bf (c)} The mean decay time $\tau_{\rm d}$ obtained from DNS  are plotted as red dots ($L=50D$) and blue squares ($L=100D$). This is the same plot as Fig.~\ref{fig:decaytime} except for the formula derived from the Fr\'echet distribution \eqref{eq:doubleexponential_after_F}, which is plotted as a red solid line ($L=50D$) and a blue solid line ($L=100D$).
}
\end{center} 
\end{figure}

We fit the Fr\'echet distribution function \eqref{eq:Frechet} to the CDF of the maximum turbulence intensity $P(h)$ \eqref{eq:cumulative} and plot it in Fig.~\ref{Fig:Frechet_1}(a). The figure indicates that $P(h)$ is approximated as 
\begin{equation}
\frac{P(h)}{P(h_{\rm th})} \simeq 
\begin{cases}
\Pi (h)   &  \quad  h \leq h_{\rm x}  \\
\ P_{\Reynul}^{\rm F}(h)/ P(h_{\rm th})  &  \quad  h > h_{\rm x}
\end{cases}
\label{eq:ph_F}
\end{equation}
with a Re-independent function $\Pi (h)$ ({\it c.f.}, Eq.\eqref{eq:ph} using the Gumbel distribution function). We observe slightly larger discrepancies in this approximation close to $h=h_{\rm x}$ than in Eq.\eqref{eq:ph}\footnote{Yet, we observe better agreement for $h>20$. See the explanation in Section~\ref{Relevance_extreme_value_distributions}.}.
For this technical reason, $h_{\rm x}$ is defined as the value that makes $P_{\Reynul}^{\rm F}(h_{\rm x})/ P(h_{\rm th})$ the same for all Re we study. These values are provided in the caption of Fig.~\ref{Fig:Frechet_1}(a) together with the values of $\Pi (h_{\rm x}) \left (\equiv P_{\Reynul}^{\rm F}(h_{\rm x})/ P(h_{\rm th}) \right )$. Following the same argument below Eq.\eqref{eq:gambel}, we then derive 
\begin{equation}
\tau_{\rm d} =  \delta t \ \Pi(h_{\rm x}) \  \exp\left [  \kappa (h_{\rm x}-\eta) ^{-\zeta} \right ].
\label{eq:doubleexponential_before_F}
\end{equation}
We next plot $\log \left [\kappa (h_{\rm x} - \eta)^{-\zeta} \right ]$ in Fig.~\ref{Fig:Frechet_1}(b) as a function of Re, which shows a linear relationship,
\begin{equation}
\log \left [ \kappa (h_{\rm x}-\eta) ^{-\zeta} \right ] \simeq \tilde a {\rm Re} + \tilde b.
\label{eq:logkappa_linear}
\end{equation}
Here the values of the parameters $\tilde a, \tilde b$ are provided in the caption of Fig.~\ref{Fig:Frechet_1}(b).
We  thus again arrive at
\begin{equation}
\tau_{\rm d}  \simeq  \delta t \ \Pi(h_{\rm x}) \  \exp\left [ \exp \left (  \tilde a {\rm Re} + \tilde b \right ) \right ],
\label{eq:doubleexponential_after_F}
\end{equation}
which is the same as \eqref{eq:tau_decay_last}. Though the parameters in this expression ($\Pi(h_x), \tilde a, \tilde b$) are slightly different from the ones in \eqref{eq:tau_decay_last}, this \eqref{eq:doubleexponential_after_F} still produces very similar lines to the ones  plotted in Fig.~\ref{fig:decaytime} as shown in Fig.~\ref{Fig:Frechet_1}(c).

\begin{table}
\begin{center}
          \caption{\label{Table:fittingGumbel} Fitting parameters $\gamma$ and $h_0$ for the Gumbel function  (\ref{eq:gambel}) used in Fig.~\ref{fig:cumumaximum_a}(b).   
          }
          \begin{tabular}{ccccc} \hline
                        & ${\rm Re}=1900 $   & ${\rm Re}=1925 $  & ${\rm Re}=1950 $  & ${\rm Re}=1975 $ 
            \\    \hline 
            $L=50D$ \\
  $\gamma $    &       0.0935183                   & 0.0943595                           & 0.0936482                   & 0.0965271
            \\  
  $h_0  $   &       21.401                      &  23.9516                                  & 26.2484                       & 27.9671 
              \\ \hline
             $L=100D$ \\
  $\gamma $    &       0.0891413                   & 0.092231                           & 0.0940843                   &   -
            \\  
  $h_0  $   &       22.6871                      &  25.211                                  & 27.4782                       &  - 
              \\ \hline
          \end{tabular}  
        \end{center}
 \end{table}

\begin{table}
\begin{center}
          \caption{\label{Table:fittingvalues} Fitting parameters determined in Fig.~\ref{fig:cumumaximum_c}(a).   
          A linear function $S \ {\rm Re} + I$ with a slope $S$ and an intercept $I$ is fitted to
          $\bar \gamma (\Reynul) \equiv \gamma (\Reynul) / \gamma(1900)$ and $\bar h_0 ({\Reynul}) \equiv h_0({\Reynul})/h_0(1900) $.  See Table \ref{Table:fittingGumbel} for the values of $ \gamma(1900), h_0(1900) $.
          }
          \begin{tabular}{ccc} \hline
                        &  Slope $S$  &  Intercept $I$ 
            \\    \hline 
           $L=50D$  \\
    $\bar \gamma$  &       0.000356                   & 0.322           
            \\  
  $\bar h_0$  &       0.00411                     & -6.80              
              \\ \hline
           $L=100D$  \\
    $\bar \gamma$  &       0.00111                   & -1.10    
            \\  
  $\bar h_0$  &       0.00422                      & -7.02                           
          \end{tabular}  
        \end{center}
 \end{table}

\bibliographystyle{jfm}
\bibliography{draft.bib}

\begin{thebibliography}{40}
\expandafter\ifx\csname natexlab\endcsname\relax\def\natexlab#1{#1}\fi
\def\au#1{#1} \def\ed#1{#1} \def\yr#1{#1}\def\at#1{#1}\def\jt#1{\textit{#1}}
  \def\bt#1{#1}\def\bvol#1{\textbf{#1}} \def\vol#1{#1} \def\pg#1{#1}
  \def\publ#1{#1}\def\arxiv#1{#1}\def\org#1{#1}\def\st#1{\textit{#1}}

\bibitem[Alexakis \& Biferale(2018)]{alexakis2018cascades}
{\sc \au{Alexakis, Alexandros} \& \au{Biferale, Luca}} \yr{2018}  \at{Cascades
  and transitions in turbulent flows}.  \jt{Physics Reports}  \bvol{767},
  \pg{1--101}.

\bibitem[Allen {\em et~al.\/}(2005)Allen, Warren \& ten Wolde]{Allen_2005}
{\sc \au{Allen, Rosalind~J.}, \au{Warren, Patrick~B.} \& \au{ten Wolde,
  Pieter~Rein}} \yr{2005}  \at{Sampling rare switching events in biochemical
  networks}.  \jt{Phys. Rev. Lett.}  \bvol{94},  \pg{018104}.

\bibitem[Avila {\em et~al.\/}(2011)Avila, Moxey, de~Lozar, Avila, Barkley \&
  Hof]{Avila192}
{\sc \au{Avila, Kerstin}, \au{Moxey, David}, \au{de~Lozar, Alberto}, \au{Avila,
  Marc}, \au{Barkley, Dwight} \& \au{Hof, Bj{\"o}rn}} \yr{2011}  \at{The onset
  of turbulence in pipe flow}.  \jt{Science}  \bvol{333}~(6039),
  \pg{192--196},  \arxiv{arXiv:
  http://science.sciencemag.org/content/333/6039/192.full.pdf}.

\bibitem[Avila {\em et~al.\/}(2010)Avila, Willis \& Hof]{avila2010transient}
{\sc \au{Avila, Marc}, \au{Willis, Ashley~P} \& \au{Hof, Bj{\"o}rn}} \yr{2010}
  \at{On the transient nature of localized pipe flow turbulence}.  \jt{Journal
  of Fluid Mechanics}  \bvol{646},  \pg{127--136}.

\bibitem[Barkley(2011)]{PhysRevE.84.016309}
{\sc \au{Barkley, Dwight}} \yr{2011}  \at{Simplifying the complexity of pipe
  flow}.  \jt{Phys. Rev. E}  \bvol{84},  \pg{016309}.

\bibitem[Benavides \& Alexakis(2017)]{benavides_alexakis_2017}
{\sc \au{Benavides, Santiago~Jose} \& \au{Alexakis, Alexandros}} \yr{2017}
  \at{Critical transitions in thin layer turbulence}.  \jt{Journal of Fluid
  Mechanics}  \bvol{822},  \pg{364--385}.

\bibitem[Bouchet {\em et~al.\/}(2019)Bouchet, Rolland \&
  Simonnet]{bouchet2018rare}
{\sc \au{Bouchet, Freddy}, \au{Rolland, Joran} \& \au{Simonnet, Eric}}
  \yr{2019}  \at{Rare event algorithm links transitions in turbulent flows with
  activated nucleations}.  \jt{Phys. Rev. Lett.}  \bvol{122},  \pg{074502}.

\bibitem[Box \& Tiao(2011)]{box2011bayesian}
{\sc \au{Box, George~EP} \& \au{Tiao, George~C}} \yr{2011} {\em Bayesian
  inference in statistical analysis\/}, ,  \vol{vol.~40}.  \publ{John Wiley \&
  Sons}.

\bibitem[Celani {\em et~al.\/}(2010)Celani, Musacchio \&
  Vincenzi]{celani2010morethantwo}
{\sc \au{Celani, Antonio}, \au{Musacchio, Stefano} \& \au{Vincenzi, Dario}}
  \yr{2010}  \at{Turbulence in more than two and less than three dimensions}.
  \jt{Phys. Rev. Lett.}  \bvol{104},  \pg{184506}.

\bibitem[C{\'e}rou \& Guyader(2007)]{cerou2007adaptive}
{\sc \au{C{\'e}rou, Fr{\'e}d{\'e}ric} \& \au{Guyader, Arnaud}} \yr{2007}
  \at{Adaptive multilevel splitting for rare event analysis}.  \jt{Stochastic
  Analysis and Applications}  \bvol{25}~(2),  \pg{417--443}.

\bibitem[Charras-Garrido \& Lezaud(2013)]{charras2013extreme}
{\sc \au{Charras-Garrido, Myriam} \& \au{Lezaud, Pascal}} \yr{2013}
  \at{Extreme value analysis: an introduction}.  \jt{Journal de la Soci\'et\'e
  Française de Statistique}  \bvol{154},  \pg{66}.

\bibitem[Chernykh \& Stepanov(2001)]{Chernykh2001}
{\sc \au{Chernykh, A.~I.} \& \au{Stepanov, M.~G.}} \yr{2001}  \at{Large
  negative velocity gradients in burgers turbulence}.  \jt{Phys. Rev. E}
  \bvol{64},  \pg{026306}.

\bibitem[Eckhardt(2009)]{Eckhardt449}
{\sc \au{Eckhardt, Bruno}} \yr{2009}  \at{Introduction. turbulence transition
  in pipe flow: 125th anniversary of the publication of
  reynolds{\textquoteright} paper}.  \jt{Philosophical Transactions of the
  Royal Society of London A: Mathematical, Physical and Engineering Sciences}
  \bvol{367}~(1888),  \pg{449--455},  \arxiv{arXiv:
  http://rsta.royalsocietypublishing.org/content/367/1888/449.full.pdf}.

\bibitem[Eckhardt {\em et~al.\/}(2007)Eckhardt, Schneider, Hof \&
  Westerweel]{doi:10.1146/annurev.fluid.39.050905.110308}
{\sc \au{Eckhardt, Bruno}, \au{Schneider, Tobias~M.}, \au{Hof, Bjorn} \&
  \au{Westerweel, Jerry}} \yr{2007}  \at{Turbulence transition in pipe flow}.
  \jt{Annual Review of Fluid Mechanics}  \bvol{39}~(1),  \pg{447--468}.

\bibitem[Fisher \& Tippett(1928)]{fisher1928limiting}
{\sc \au{Fisher, Ronald~Aylmer} \& \au{Tippett, Leonard Henry~Caleb}} \yr{1928}
  Limiting forms of the frequency distribution of the largest or smallest
  member of a sample.  \bt{In {\em Mathematical Proceedings of the Cambridge
  Philosophical Society\/}}, ,  \vol{vol.~24},  \pg{pp. 180--190}. Cambridge
  University Press.

\bibitem[Giardin\`a {\em et~al.\/}(2006)Giardin\`a, Kurchan \&
  Peliti]{Giadina_2006}
{\sc \au{Giardin\`a, Cristian}, \au{Kurchan, Jorge} \& \au{Peliti, Luca}}
  \yr{2006}  \at{Direct evaluation of large-deviation functions}.  \jt{Phys.
  Rev. Lett.}  \bvol{96},  \pg{120603}.

\bibitem[Goldenfeld {\em et~al.\/}(2010)Goldenfeld, Guttenberg \&
  Gioia]{PhysRevE.81.035304}
{\sc \au{Goldenfeld, Nigel}, \au{Guttenberg, Nicholas} \& \au{Gioia, Gustavo}}
  \yr{2010}  \at{Extreme fluctuations and the finite lifetime of the turbulent
  state}.  \jt{Phys. Rev. E}  \bvol{81},  \pg{035304(R)}.

\bibitem[Goldenfeld \& Shih(2017)]{Goldenfeld2017}
{\sc \au{Goldenfeld, Nigel} \& \au{Shih, Hong-Yan}} \yr{2017}  \at{Turbulence
  as a problem in non-equilibrium statistical mechanics}.  \jt{Journal of
  Statistical Physics}  \bvol{167}~(3),  \pg{575--594}.

\bibitem[Grafke {\em et~al.\/}(2015{\natexlab{{\em a\/}}})Grafke, Grauer \&
  Sch{\"a}fer]{Grafke_2015}
{\sc \au{Grafke, Tobias}, \au{Grauer, Rainer} \& \au{Sch{\"a}fer, Tobias}}
  \yr{2015{\natexlab{{\em a\/}}}}  \at{The instanton method and its numerical
  implementation in fluid mechanics}.  \jt{Journal of Physics A: Mathematical
  and Theoretical}  \bvol{48}~(33),  \pg{333001}.

\bibitem[Grafke {\em et~al.\/}(2015{\natexlab{{\em b\/}}})Grafke, Grauer,
  Sch{\"a}fer \& Vanden-Eijnden]{Grafke_2015_2}
{\sc \au{Grafke, T.}, \au{Grauer, R.}, \au{Sch{\"a}fer, T.} \&
  \au{Vanden-Eijnden, E.}} \yr{2015{\natexlab{{\em b\/}}}}  \at{Relevance of
  instantons in burgers turbulence}.  \jt{{EPL} (Europhysics Letters)}
  \bvol{109}~(3),  \pg{34003}.

\bibitem[Grigorio {\em et~al.\/}(2017)Grigorio, Bouchet, Pereira \&
  Chevillard]{Grigorio_2017}
{\sc \au{Grigorio, L~S}, \au{Bouchet, F}, \au{Pereira, R~M} \& \au{Chevillard,
  L}} \yr{2017}  \at{Instantons in a lagrangian model of turbulence}.
  \jt{Journal of Physics A: Mathematical and Theoretical}  \bvol{50}~(5),
  \pg{055501}.

\bibitem[Gumbel(1935)]{gumbel1935valeurs}
{\sc \au{Gumbel, Emil~Julius}} \yr{1935}  \at{Les valeurs extr{\^e}mes des
  distributions statistiques}.  \jt{Ann. Inst. Henri Poincar{\'e}}
  \bvol{5}~(2),  \pg{115--158}.

\bibitem[Heymann \& Vanden-Eijnden(2008)]{Heymann2008}
{\sc \au{Heymann, Matthias} \& \au{Vanden-Eijnden, Eric}} \yr{2008}
  \at{Pathways of maximum likelihood for rare events in nonequilibrium systems:
  Application to nucleation in the presence of shear}.  \jt{Phys. Rev. Lett.}
  \bvol{100},  \pg{140601}.

\bibitem[Hof {\em et~al.\/}(2008)Hof, de~Lozar, Kuik \&
  Westerweel]{PhysRevLett.101.214501}
{\sc \au{Hof, Bj\"orn}, \au{de~Lozar, Alberto}, \au{Kuik, Dirk~Jan} \&
  \au{Westerweel, Jerry}} \yr{2008}  \at{Repeller or attractor? selecting the
  dynamical model for the onset of turbulence in pipe flow}.  \jt{Phys. Rev.
  Lett.}  \bvol{101},  \pg{214501}.

\bibitem[Hof {\em et~al.\/}(2006)Hof, Westerweel, Schneider \&
  Eckhardt]{hof2006finite}
{\sc \au{Hof, Bj{\"o}rn}, \au{Westerweel, Jerry}, \au{Schneider, Tobias~M} \&
  \au{Eckhardt, Bruno}} \yr{2006}  \at{Finite lifetime of turbulence in shear
  flows}.  \jt{Nature}  \bvol{443}~(7107),  \pg{59--62}.

\bibitem[van Kan {\em et~al.\/}(2019)van Kan, Nemoto \&
  Alexakis]{van_kan_nemoto_alexakis_2019}
{\sc \au{van Kan, Adrian}, \au{Nemoto, Takahiro} \& \au{Alexakis, Alexandros}}
  \yr{2019}  \at{Rare transitions to thin-layer turbulent condensates}.
  \jt{Journal of Fluid Mechanics}  \bvol{878},  \pg{356--369}.

\bibitem[Kuik {\em et~al.\/}(2010)Kuik, Poelma \&
  Westerweel]{kuik2010quantitative}
{\sc \au{Kuik, Dirk~Jan}, \au{Poelma, C.} \& \au{Westerweel, Jerry}} \yr{2010}
  \at{Quantitative measurement of the lifetime of localized turbulence in pipe
  flow}.  \jt{Journal of fluid mechanics}  \bvol{645},  \pg{529--539}.

\bibitem[Lestang {\em et~al.\/}(2018)Lestang, Ragone, Br{\'{e}}hier, Herbert \&
  Bouchet]{Lestang_2018}
{\sc \au{Lestang, Thibault}, \au{Ragone, Francesco}, \au{Br{\'{e}}hier,
  Charles-Edouard}, \au{Herbert, Corentin} \& \au{Bouchet, Freddy}} \yr{2018}
  \at{Computing return times or return periods with rare event algorithms}.
  \jt{Journal of Statistical Mechanics: Theory and Experiment}
  \bvol{2018}~(4),  \pg{043213}.

\bibitem[de~Lozar \& Hof(2009)]{de_Lozar589}
{\sc \au{de~Lozar, Alberto} \& \au{Hof, Bj{\"o}rn}} \yr{2009}  \at{An
  experimental study of the decay of turbulent puffs in pipe flow}.
  \jt{Philosophical Transactions of the Royal Society of London A:
  Mathematical, Physical and Engineering Sciences}  \bvol{367}~(1888),
  \pg{589--599},  \arxiv{arXiv:
  http://rsta.royalsocietypublishing.org/content/367/1888/589.full.pdf}.

\bibitem[de~Lozar {\em et~al.\/}(2012)de~Lozar, Mellibovsky, Avila \&
  Hof]{PhysRevLett.108.214502}
{\sc \au{de~Lozar, A.}, \au{Mellibovsky, F.}, \au{Avila, M.} \& \au{Hof, B.}}
  \yr{2012}  \at{Edge state in pipe flow experiments}.  \jt{Phys. Rev. Lett.}
  \bvol{108},  \pg{214502}.

\bibitem[Musacchio \& Boffetta(2017)]{musacchio2017split}
{\sc \au{Musacchio, Stefano} \& \au{Boffetta, Guido}} \yr{2017}  \at{Split
  energy cascade in turbulent thin fluid layers}.  \jt{Physics of Fluids}
  \bvol{29}~(11),  \pg{111106}.

\bibitem[Nemoto \& Alexakis(2018)]{PhysRevE.97.022207}
{\sc \au{Nemoto, Takahiro} \& \au{Alexakis, Alexandros}} \yr{2018}  \at{Method
  to measure efficiently rare fluctuations of turbulence intensity for
  turbulent-laminar transitions in pipe flows}.  \jt{Phys. Rev. E}  \bvol{97},
  \pg{022207}.

\bibitem[Nemoto {\em et~al.\/}(2016)Nemoto, Bouchet, Jack \&
  Lecomte]{Nemoto_2016}
{\sc \au{Nemoto, Takahiro}, \au{Bouchet, Freddy}, \au{Jack, Robert~L.} \&
  \au{Lecomte, Vivien}} \yr{2016}  \at{Population-dynamics method with a
  multicanonical feedback control}.  \jt{Phys. Rev. E}  \bvol{93},
  \pg{062123}.

\bibitem[Reynolds(1883)]{reynolds1883experimental}
{\sc \au{Reynolds, Osborne}} \yr{1883}  \at{An experimental investigation of
  the circumstances which determine whether the motion of water shall be direct
  or sinuous, and of the law of resistance in parallel channels}.
  \jt{Proceedings of the royal society of London}  \bvol{35}~(224--226),
  \pg{84--99}.

\bibitem[Shimizu {\em et~al.\/}(2019)Shimizu, Kanazawa \&
  Kawahara]{Shimizu_2019}
{\sc \au{Shimizu, Masaki}, \au{Kanazawa, Takahiro} \& \au{Kawahara, Genta}}
  \yr{2019}  \at{Exponential growth of lifetime of localized turbulence with
  its extent in channel flow}.  \jt{Fluid Dynamics Research}  \bvol{51}~(1),
  \pg{011404}.

\bibitem[Smith {\em et~al.\/}(1996)Smith, Chasnov \&
  Waleffe]{smith1996crossover}
{\sc \au{Smith, Leslie~M}, \au{Chasnov, Jeffrey~R} \& \au{Waleffe, Fabian}}
  \yr{1996}  \at{Crossover from two-to three-dimensional turbulence}.
  \jt{Physical review letters}  \bvol{77}~(12),  \pg{2467}.

\bibitem[Tailleur \& Kurchan(2007)]{tailleur2007probing}
{\sc \au{Tailleur, Julien} \& \au{Kurchan, Jorge}} \yr{2007}  \at{Probing rare
  physical trajectories with lyapunov weighted dynamics}.  \jt{Nature Physics}
  \bvol{3}~(3),  \pg{203}.

\bibitem[Teo {\em et~al.\/}(2016)Teo, Mayne, Schulten \& Leli\`evre]{Teo_2016}
{\sc \au{Teo, Ivan}, \au{Mayne, Christopher~G.}, \au{Schulten, Klaus} \&
  \au{Leli\`evre, Tony}} \yr{2016}  \at{Adaptive multilevel splitting method
  for molecular dynamics calculation of benzamidine-trypsin dissociation time}.
   \jt{Journal of Chemical Theory and Computation}  \bvol{12}~(6),
  \pg{2983--2989}, pMID: 27159059.

\bibitem[Willis(2017)]{WILLIS2017124}
{\sc \au{Willis, Ashley~P.}} \yr{2017}  \at{The openpipeflow navier-stokes
  solver}.  \jt{SoftwareX}  \bvol{6},  \pg{124 -- 127}.

\bibitem[Wygnanski \& Champagne(1973)]{wygnanski1973transition}
{\sc \au{Wygnanski, I.~J.} \& \au{Champagne, F.~H.}} \yr{1973}  \at{On
  transition in a pipe. part 1. the origin of puffs and slugs and the flow in a
  turbulent slug}.  \jt{Journal of Fluid Mechanics}  \bvol{59}~(02),
  \pg{281--335}.

\end{thebibliography}

\end{document}